\documentclass[11pt]{article}

\usepackage{graphicx} 
\usepackage[margin=1in]{geometry} 
\graphicspath{{Figures/}} 
\usepackage{amsmath} 
\usepackage{amssymb} 
\usepackage{multirow} 
\usepackage{booktabs} 
\usepackage{color} 
\usepackage{indentfirst} 
\usepackage{caption, subcaption} 
\usepackage{bm} 
\usepackage{siunitx} 
\usepackage[noblocks]{authblk} 

\newcommand{\vect}[1]{\bm{#1}}

\begin{document}
	
\newcommand{\beginsupplement}{%
		\setcounter{table}{0}
		\setcounter{figure}{0}
		\setcounter{section}{0}
		\setcounter{subsection}{0}
		\setcounter{equation}{0}
		\setcounter{paragraph}{0}
		\setcounter{page}{1}
		\renewcommand{\thetable}{S\arabic{table}}%
		\renewcommand{\thefigure}{S\arabic{figure}}%
		\renewcommand{\thesection}{S\arabic{section}}%
		\renewcommand{\thesubsection}{S\arabic{section}.\arabic{subsection}}%
		\renewcommand{\theequation}{S\arabic{equation}}%
		\renewcommand{\theparagraph}{S\arabic{paragraph}}%
		\renewcommand{\thepage}{S\arabic{page}}%
		\renewcommand{\theHtable}{S\the\value{table}}%
		\renewcommand{\theHfigure}{S\the\value{figure}}%
		\renewcommand{\theHsection}{S\the\value{section}}%
		\renewcommand{\theHsubsection}{S\the\value{subsection}}%
		\renewcommand{\theHequation}{S\the\value{equation}}%
		\renewcommand{\theHparagraph}{S\the\value{paragraph}}%
}
	
\title{Multi-point Scattering Measurements for Effective Property Extraction from Metamaterials with Skin Effects}
	\author[1]{Joshua Morris}
	\author[1,2]{Alireza Amirkhizi}
	\affil[1]{University of Massachusetts Lowell, Lowell, MA 01854}
	\affil[2]{alireza\_amirkhizi@uml.edu}
	\date{\today}
	\maketitle

	\renewenvironment{abstract}
	{\begin{quote}
			\noindent \rule{\linewidth}{.5pt}\par{\bfseries \abstractname.}}
		{\medskip\noindent \rule{\linewidth}{.5pt}
	\end{quote}}
\begin{abstract}
	Scattering experiments can be leveraged to extract the effective properties of a heterogeneous metamaterial slab based on multi-point measurements in surrounding media. In this technique, two measurements are made in the ambient media on each side of a finite thickness micro-structured slab to decompose incoming and outgoing waves. The method is applied to an example with locally resonant micro-structured inclusions while paying close attention to parameters that influence the extracted material parameters. It is observed that the extracted overall parameters converge to limiting values when the number of unit cells across the slab thickness or ambient media modulus are increased. Dependence of extracted material parameters on the number of cells through thickness or ambient media properties are attributed to the different response of boundary (skin) and interior cells. A method is presented which represents a finite array with different effective properties for the skin regions vs. the interior regions with great success in reproducing the scattering for different slab thicknesses, and through which the interior regions properties become independent of ambient media properties. In stop bands with lower material loss, challenges arise that are associated with extremely small transmission. The assumption of continuity in transmission phase is enforceable to remove phase ambiguity, although in certain cases (associated with nearly lossless specimens) this assumption appears to fail. In such cases, interference from coupled shear modes may lead to apparent higher longitudinal transmission within the stop band. This work provides a proof of concept for the development of an experimental methodology capable of extracting the effective properties and dispersion behavior of heterogeneous mechanical metamaterials without any knowledge of the internal fields.
\end{abstract}

\maketitle

\textbf{Highlights.}
	\begin{itemize}
        \item Multi-point measurements enable property extraction from phase-resolved scattering.
        \item Skin region properties differ from interior and is affected by the ambient media.
        \item Closed form solutions are derived for skin and interior regions properties.
        \item Coupled shear modes permit energy transmission with minimal cell asymmetry.
        
    \end{itemize}
		
{\bf Keywords.} mechanical metamaterials, scattering analysis, overall constitutive parameters, transmission/reflection experiment, wave propagation, boundary layer
	
\section{Introduction}

The effective properties of micro-structured materials are constitutive parameters which when applied to a continuum specimen of the same overall geometry, would reproduce the behavior of the original heterogeneous system. For dynamic mechanical metamaterials this behavior may be considered as either proper averages of fields inside the domain, or scattering response of a finite specimen. A number of such methods are introduced in the past including field averaging/integration \cite{amirkhizi_homogenization_2017,zhu_effective_2012,willis_effective_2011,norris_analytical_2012,srivastava_overall_2012}, multiple scattering theory \cite{brunet_soft_2015,mei_theory_2003}, plane wave expansion method \cite{kushwaha_acoustic_1993,montero_de_espinosa_ultrasonic_1998}, variational method \cite{goffaux_two-dimensional_2003}, finite difference time domain method \cite{tanaka_band_2000}, or extraction of the effective properties from the scattering parameters or impedance \cite{yang_homogenization_2014,amirkhizi_overall_2018}. Field averaging/integration works particularly well for systems where the periodic unit cell is much smaller than the excitation wave length and the metamaterial can be adequately homogenized \cite{amirkhizi_homogenization_2017,aghighi_low-frequency_2019}, but would require the knowledge of of full field quantities inside a specimen, which is not feasible in experiments. On the other hands scattering-based methods rely only on the measurable quantities as the specimen interacts with ambient domains. Therefore, experimental characterization of heterogeneous media at acoustic or ultrasonic frequencies is typically performed by analyzing amplitudes and extracting the macro-scale properties from the scattering coefficients. Impedance tubes \cite{lu_lightweight_2016,sui_lightweight_2015,yang_acoustic_2010}, contact transducer \cite{brunet_soft_2015}, or non-contact transducer \cite{fokin_method_2007} experiments are common methods for collecting scattering data. Contact and non-contact techniques often include one measurement point or an amplitude average of a time signal. Phase velocity information has been extracted by conducting two tests with different sample lengths and comparing the phase angles of the FFT values for the two signals \cite{lee_composite_2010,fang_ultrasonic_2006}. Impedance tubes have incorporated up to four measurement points and often utilize phase data to separate overlapping wave forms \cite{yang_membrane-type_2008,langfeldt_membrane-type_2016}. The challenges specific to impedance tubes include instability at higher frequencies ($>$ \SI{7}{kHz}), use of fluid ambient media not allowing to measure shear waves, and inability to process samples with intricate porosity \cite{koruk_assessment_2014,hiremath_overview_2021}. Air can be used in the impedance tubes \cite{ghaffarivardavagh_ultra-open_2019,romero-garcia_design_2020}, however the frequency range becomes greatly limited. Many of these challenges can be overcome by using a solid material as the ambient medium for the measurement of scattered (transmitted and reflected) longitudinal and shear waves. For each of these techniques, an assumption of the material's apparent or effective dynamic density (typically simplified to be equal to its static value and constant over the frequency range) is required in order to evaluate further properties \cite{koruk_assessment_2014}. This becomes problematic when characterizing metamaterial dispersion properties, potentially overlooking unique qualities in both effective density and effective modulus \cite{amirkhizi_homogenization_2017}. Furthermore, adoption of these methods to a compact laboratory experimental characterization technique requires the ability to use measurements that may include reflections from boundaries of the experimental setup. The method originally introduced in \cite{aghighi_low-frequency_2019} and expanded here utilizes multiple measurement points on each side of the specimen to identify the reflections from the boundaries of a finite sized experimental domain. All available information regarding the behavior of the specimen are collected in this manner and therefore do not require knowledge of interior fields or features. The overall or apparent constitutive parameters, therefore, can be determined using this information and only this information. The focus of this paper is to explore the robustness and challenges of this characterization method, developed within a simulated environment. 
	
Conducting reliable amplitude and phase measurements is a challenging task and many parameters could affect the application of this technique. Physical considerations related to the design of a future experiment include: the material properties of the ambient media, distance of the measurement locations to the specimen, thickness of the specimen, and what field measurement is performed (e.g. strain, particle velocity, or traction). For the sake of brevity, within this paper only the parameters that do influence the apparent extracted parameters are presented in detail, while it is stated that the remaining physical parameters did not affect the results. Key factors that would lead to the most accurate solution and aid with the design of a successful characterization experiment are identified, two of which deserve further emphasis here. The scattering response of a specimen is not only related to the material parameters of the specimen, but also on those of its surroundings and the geometry of the specimen. For homogeneous media, overall properties can be determined that would describe the response of the test article in all such variations. However, perfect homogenization of micro-structured media faces a limitation related to the difference in dynamic response for boundary (skin) cells and interior cells \cite{srivastava_limit_2014}. Techniques to adjust for boundary dependencies have included running multiple arrays of varying size to extract parameters for individual cells \cite{li_hou_advanced_2008}, treating the outer layer to be evanescent \cite{srivastava_evanescent_2017}, or enforcing interface continuity with Betti-Rayleigh reciprocity theorem \cite{mokhtari_scattering_2020}. A method is proposed here to ascribe different apparent properties to skin versus the interior regions, which is successful in reproducing the scattering behavior independent of the thickness of the array, and for which the interior region properties are not affected by the ambient media properties.

The structure of this paper is as follows. First, standard eigenfrequency analysis is used to determine the band structure of an infinitely periodic array of specific low frequency resonant metamaterial unit cells. The quasi-static compressive response of a single cell is also determined to provide a low frequency limit and basis for comparison. Next a model is presented to determine the scattering of a finite thickness specimen, along with the analysis technique required for extracting the apparent effective properties (based on 2-point measurements of the harmonic fields on either side of the finite thickness specimen). Of all the physical parameters discussed, only the numerical parametric sweeps that show the influence of the ambient media choice and number of metamaterial cells through thickness are presented. The physical reasons why these parameters would affect the results will be discussed. The influence of ambient media on the boundary cell properties is presented next and two sets of properties (skin vs. interior) are shown which can reproduce scattering of any thickness of specimen. Additionally, the loss effect is understood and utilized for cases where the very low transmission in stop bands complicate the process based on scattering (transmission/reflection) measurements. Consequently, the penultimate section highlights the challenges specific to the stop bands with very low loss, especially when there is potential for coupled and interfering modes contributing to the transmission. Finally, the observations are summarized and used to provide the basis for an experiment capable of replicating this characterization technique in a physical laboratory.
	
\section{Eigenfrequency Analysis and Quasi-Static Mechanical Properties}
	
\begin{figure}[!ht]
		\centering
		\begin{subfigure}{0.28\linewidth}
			\includegraphics[scale=0.15]{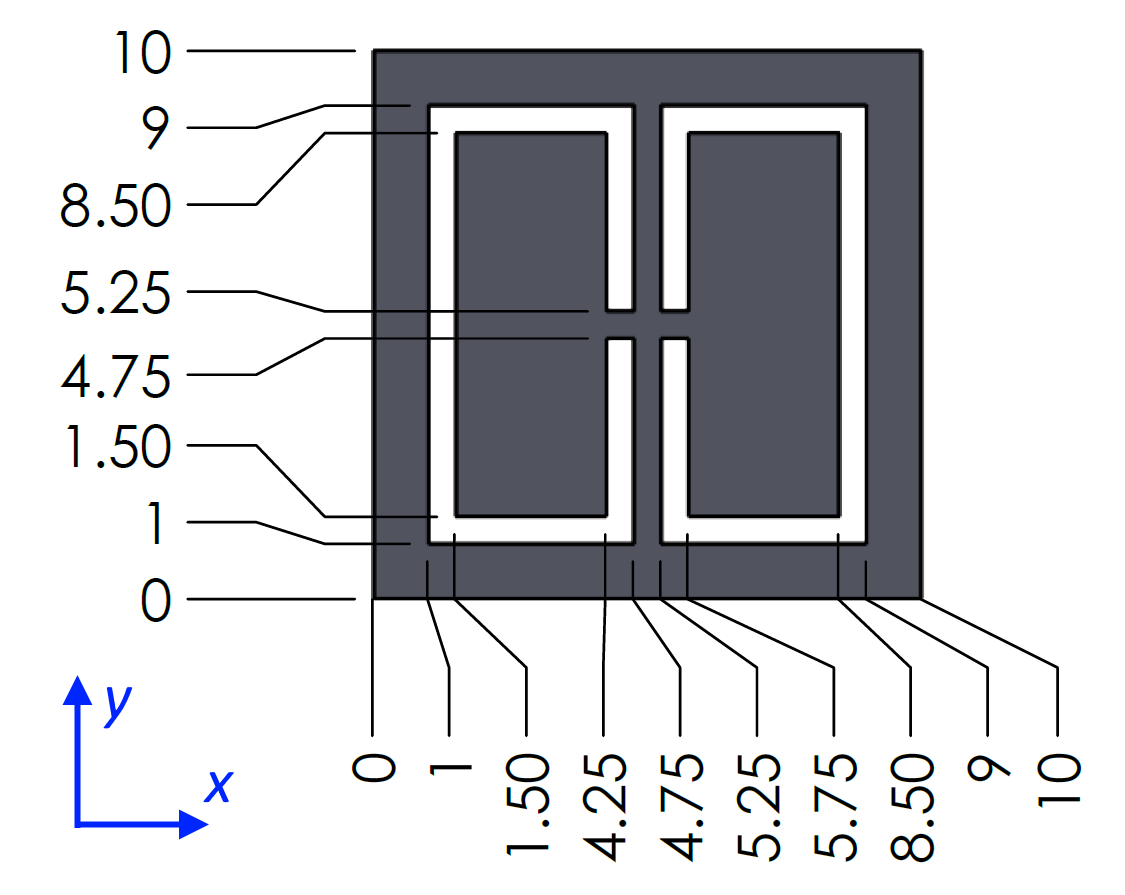}
			\caption{ }
			\label{geometry_3khzH}
		\end{subfigure}
		\begin{subfigure}{0.35\linewidth}
			\includegraphics[scale=0.22]{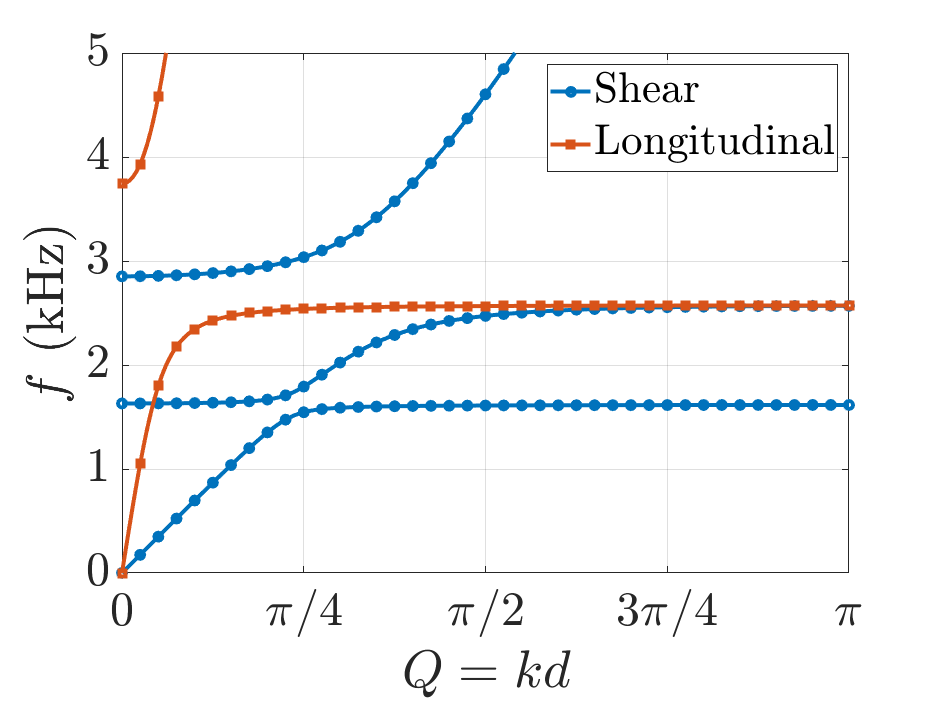}
			\caption{}
			\label{H_band} 
		\end{subfigure}
		\begin{subfigure}{0.35\linewidth}
			\includegraphics[scale=0.22]{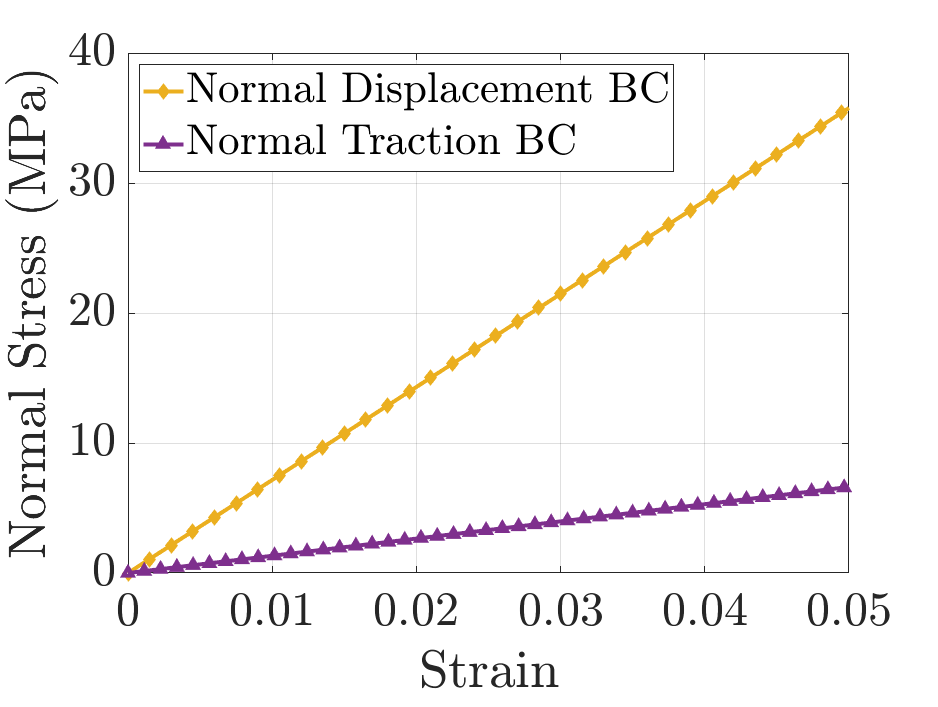}
			\caption{}
			\label{H_moduli}
		\end{subfigure}
		\caption{(\subref{geometry_3khzH}) The geometry of the \SI{10}{mm} H metamaterial cell. (\subref{H_band}) The band structure calculated using eigenfrequency analysis. (\subref{H_moduli}) Quasi-static compression model for the cell with normal displacement and traction boundary conditions.}
		\label{H_eigen} 
	\end{figure}

Eigenfrequency analysis \cite{aghighi_low-frequency_2019} is the standard approach for calculation of the band structure of infinitely periodic micro-structures, which collects the wave dispersion information in a compact form. The procedure requires solving the dynamic eigenvalue problem of a repeating unit cell with Floquet boundary conditions and recording the frequencies and mode shapes associated with each value of the wave vector. In this work, square frame with lattice spacing of $a_x = a_y =$ \SI{10}{mm} and a H-shaped inclusion with resonance frequency around \SI{3}{kHz} is analyzed along the $x$-axis as shown in Figure~\ref{geometry_3khzH}. The band structure and quasi-static properties of this cell were presented in \cite{aghighi_low-frequency_2019}, from which the relevant results are reproduced in Figures~\ref{H_band} and \ref{H_moduli}. The results were generated using eigenfrequency and structural simulations in COMSOL Multiphysics. The isotropic modulus, Poisson's ratio, mass density, and loss factor of the resin are $E_{res}=$ \SI{2.762}{GPa}, $\nu_{res}=$ \num{0.35}, $\rho_{res}=$ \SI{1161}{kg/m^3}, and $\eta_{res}=0.02$, respectively. The inclusion creates one full and two partial in-plane band gaps below \SI{5}{kHz} (longitudinal from \SIrange{2573}{3749}{Hz}, highlighted in Figure~\ref{H_band}, and shear from \SIrange{2569}{2852}{Hz} and from \SIrange{1616}{1630}{Hz}). The figure of merit for this low-frequency resonant metamaterial array, defined as the ratio of the longitudinal wavelength in the resin to lattice spacing calculated at the center of stop band is $\lambda_{res}/a_x = 61.6$. The anti-plane shear modes are omitted here and in the 2D analysis.
	
The average mass density of the unit cell is ($\rho_{0} =$ \SI{917}{kg/m^{3}}), which is simply calculated based on the proportion of the area of the square unit cell that is filled with resin. The overall quasi-static response of the unit cell is shown in Figure~\ref{H_moduli}. Heuristic bounds for the cell compressive properties can be formed using normal displacement and normal traction boundary conditions, where the tangential tractions are zero for both cases. The average slopes of the stress-strain curves were calculated to provide the longitudinal modulus ($L_{qs}$). The long wavelength approximations of the wave speed ($c_{qs}$) and effective impedance ($Z_{qs}$) can then be determined from 
	\begin{equation} \label{eq:speed}
	  c_{qs}= \sqrt{\frac{L_{qs}}{\rho_{0}}},
	\end{equation}
	\begin{equation} \label{eq:imp}
	  Z_{qs}=\sqrt{\rho_{0} L_{qs}}.
	\end{equation}
All of the quasi-static results extracted from these simulations are collected in Table~\ref{table_moduli}. The wave speed (phase velocity) determined from the band structure as frequency approaches $0$ Hz is a nearly exact match with the value determined from the quasi-static simulations using displacement BC and the average mass density value, $\rho_0$. 

	\begin{table}[!ht]
	\centering
	\caption{Quasi-static properties of a 2D periodic array with unit cell shown in Figure~\ref{geometry_3khzH}, including average mass density ($\rho_{0}$), quasi-static longitudinal modulus ($L_{qs}$), and long wavelength limit wave speed ($c_{qs}$), and effective impedance ($Z_{qs}$).}
	\label{table_moduli} 
	\begin{tabular}{cccccccc}
		\hline
		  \multirow{3}{*}{Cell Geometry}   &       & \multicolumn{3}{c}{Normal Displacement BC} & \multicolumn{3}{c}{Normal Traction BC} \\
		\cmidrule(lr){3-5} \cmidrule(lr){6-8} & $\rho_{0}$ & $L_{qs}$ & $c_{qs}$ &  $Z_{qs}$  & $L_{qs}$ & $c_{qs}$ & $Z_{qs}$ \\
		                   & (\si{kg/m^{3}}) & (\si{MPa}) & (\si{m/s}) &  (\si{MRayl})  & (\si{MPa}) & (\si{m/s}) & (\si{MRayl}) \\ \hline
		      \SI{10}{mm} H-Design      &   917   &  715  &  883  &   0.81   &  132  &  380  &  0.35
	\end{tabular}
	\end{table}
	
\section{Derivation of the Overall Constitutive Parameters Based on Scattering} \label{scattering_analysis}
	
The effective properties extraction technique introduced in \cite{aghighi_low-frequency_2019} is used here to study and analyze metamaterial specimens of finite thickness. The process for converting numerically calculated values of velocity or traction (equivalently strain in an experiment) as functions of frequency into effective properties is summarized with attention to the specific response of the \SI{10}{mm} H unit cell model.
	
\begin{figure}[!ht]
		\centering
		\includegraphics [scale=0.35]{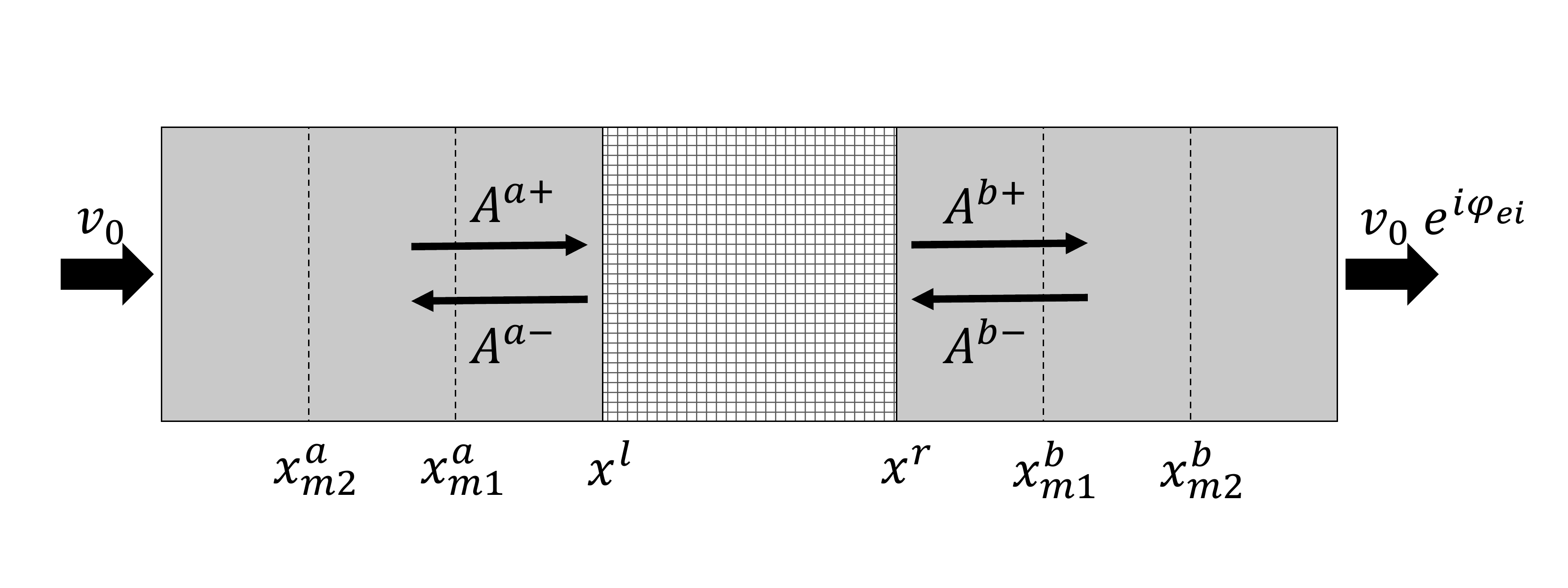}
		\caption{Notation for measurement domains ($\alpha = a, \, b$), measurement locations ($x^{\alpha}_{m1, \, m2}$), prescribed boundary conditions (in this case, velocities $v_{0}, \, v_{0} e^{i\varphi_{ei}}$), and traveling waves ($A^{\alpha\pm}$) used in the scattering property extraction technique. The inner, checkered region is an unknown specimen while the outer, gray regions are known ambient media.}
		\label{notation} 
	\end{figure}

The scattering parameters in normal incidence are defined based on the forward and backward traveling wave components $A^{\alpha\pm}$, where $\alpha$ represents the domains to the left ($a$) or right ($b$) of the specimen. The ``outgoing'' waves are linearly dependent upon the ``incoming'' waves:
  \begin{equation} \label{eq:sm}
	\begin{pmatrix}
	A^{a-} \\
	A^{b+} 
	\end{pmatrix}
	=
	\begin{pmatrix}
	\mathsf{S}_{aa} & \mathsf{S}_{ab} \\
	\mathsf{S}_{ba} & \mathsf{S}_{bb}
	\end{pmatrix}
	\begin{pmatrix}
	A^{a+} \\
	A^{b-}
	\end{pmatrix}.
  \end{equation}
Assuming full knowledge of all four waves, a single ``experiment'' provides two linear equations for the four unknown components of the scattering matrix. Therefore, two linearly independent experiments would fully determine the scattering matrix. For a typical case using a single measurement point on each side (e.g. $x_{m1}^{\alpha}$), the two waves in $a$ or $b$ domains are overlapping and the distinct directional components of the traveling waves cannot be isolated. This could be resolved in time-domain modeling or experimentation by using short pulses and long enough measurement domains to create a time delay sufficient to separate the signals. However, for long wavelengths (low frequencies) the required dimensions become infeasible and too expensive to replicate in a physical experiment. Alternatively, using a second point of measurement in each domain ($x_{m2}^{\alpha}$) can provide enough information to separate the incoming and outgoing waves, provided that the boundary conditions on the measurement domain opposite to the specimen are linear in nature. Here, the two required numerical experiments are performed in frequency domain (hence considering a harmonic steady state). The boundary conditions on the two ends of media $a$ and $b$ are harmonically prescribed velocity in the normal direction. In all cases used here, it is assumed the amplitudes of the prescribed velocities are the same at all points on both boundaries, though this is not a requirement. The two numerical experiments ($ei=e1, \, e2$) are identical except for the phase difference between the two boundaries (i.e. $\varphi_{e1}, \, \varphi_{e2}$). The notation used in the analysis is clarified in Figure~\ref{notation}, in which the specimen between domains $a$ and $b$ is a completely unknown heterogeneous metamaterial and all measurements are made in the known surrounding media.
	
Four complex amplitude measurements, $V$ (e.g. normal velocity $V = v_x$, normal stress component $V = \sigma_{xx}$, or in the case of experiment, normal strain $V = \varepsilon_{xx}$ integrated over a plane at measurement point $x^\alpha_{mi}$), were made in each of the two simulations. The forward and backward traveling wave amplitudes at locations $x^\alpha$ are then calculated based on the wave numbers in ambient media, $k^{\alpha\pm}$, using (based on $e^{i(\omega t - \vect{k}.\vect{x})}$ phasor convention)
  \begin{equation}
	\begin{pmatrix}
	A_{e1}^{\alpha+} & A_{e2}^{\alpha+} \\
	A_{e1}^{\alpha-} & A_{e2}^{\alpha-}
	\end{pmatrix}
	=
	(\vect{\gamma}^\alpha(x^\alpha, x^\alpha_{m1}, x^\alpha_{m2}))^{-1}
	\begin{pmatrix}
	V^{\alpha}_{e1}(x_{m1}^{\alpha}) & V^{\alpha}_{e2}(x_{m1}^{\alpha}) \\
	V^{\alpha}_{e1}(x_{m2}^{\alpha}) & V^{\alpha}_{e2}(x_{m2}^{\alpha})
	\end{pmatrix},
	\label{Eq_phase} 
	\end{equation}
where
  \begin{equation}
	\vect{\gamma}^\alpha = 
	\begin{pmatrix}
	e^{-ik^{\alpha+}(x^{\alpha}_{m_1}-x^{\alpha})} & e^{-ik^{\alpha-}(x^{\alpha}_{m_1}-x^{\alpha})} \\
	e^{-ik^{\alpha+}(x^{\alpha}_{m_2}-x^{\alpha})} & e^{-ik^{\alpha-}(x^{\alpha}_{m_2}-x^{\alpha})} 
	\end{pmatrix}.
	\label{gamma_matrix}
  \end{equation}
Of course for reciprocal ambient media $k^{\alpha-} = - k^{\alpha+} = -k^\alpha$, which will be used going forward. Each column is independent in this calculation. $x^\alpha$ is the position of the ``de-embedding'' for each domain. At $x^\alpha$ both incoming and outgoing waves share a de-embedding position, however, such a choice is not a necessity and all following calculation may also be performed with different $x^{\alpha\pm}$ if necessary, including the translation to new positions using the $\vect{\delta}$ matrix below. 
	
A state vector consisting of relevant particle velocity ($v$) and stress ($\sigma$) components at any location ($x$) may now be written as 
	\begin{equation}
	\begin{pmatrix}
	v_{ei}(x) \\
	\sigma_{ei}(x)
	\end{pmatrix}
	=
	\vect{\zeta}^{\alpha} \vect{\delta}^{\alpha}(x, x^{\alpha})
	\begin{pmatrix}
	A^{\alpha+}_{ei} \\
	A^{\alpha-}_{ei}
	\end{pmatrix}, 
	\end{equation}
where $\delta$ is a translation operator
	\begin{equation}
	\vect{\delta}^{\alpha}(x, x^\alpha) = 
	\begin{pmatrix}
	e^{-ik^{\alpha}(x-x^{\alpha})} & 0 \\
	0 & e^{ik^{\alpha}(x-x^{\alpha})} 
	\end{pmatrix}.
	\end{equation}
A simplifying choice is to de-embed to the interfaces of the sample, i.e. $x^a = x^l$ and $x^b = x^r$, where superscripts $l$ and $r$ represent the left and right interfaces of the specimen. In this case, evaluations of the state vectors at the interfaces (as is done for transfer matrix method) gives $\vect{\delta}^\alpha(x^l, x^a) = \vect{\delta}^\alpha(x^r, x^b) = I_2$, the identity matrix. The matrix $\vect{\zeta}^\alpha$ depends on the choice of measurement field $V$ and  the material properties of the measurement domains. For example,
	\begin{equation} \label{eq:Z_vel}
	\vect{\zeta}^{\alpha} = 
	\begin{pmatrix}
	1 & 1 \\
	-Z^\alpha & Z^\alpha
	\end{pmatrix},
	\end{equation}
for velocity measurements and
	\begin{equation} \label{eq:Z_stress}
	\vect{\zeta}^{\alpha} = 
	\begin{pmatrix}
	-1/Z^\alpha & 1/Z^\alpha \\
	1 & 1
	\end{pmatrix},
	\end{equation}
for stress or pressure measurements. Here $Z^\alpha$ represents the impedance of the ambient media. In experiments, strain measurements are simply proportional to stress, though care has to be taken if the measurement domains are different from each other. 
	
The transfer matrix of the test specimen can now be calculated by considering the states at the interfaces associated with the two independent ``experiments''. There is enough information to resolve the 4 unknown components of the transfer matrix. Since
	\begin{equation} \label{eq:TM}
	\begin{pmatrix}
	v^{r} \\
	\sigma^{r}
	\end{pmatrix}
	=
	\mathsf{T}
	\begin{pmatrix}
	v^{l} \\
	\sigma^{l}
	\end{pmatrix},
	\end{equation}
holds for all possible states, any two linearly independent experiments can be used to determine the transfer matrix 
	\begin{equation}
	\mathsf{T} = 
	\begin{pmatrix}
	\mathsf{T}_{11} & \mathsf{T}_{12} \\
	\mathsf{T}_{21} & \mathsf{T}_{22}
	\end{pmatrix}
	=
	\begin{pmatrix}
	v_{e1}^{r} & v_{e2}^{r} \\
	\sigma_{e1}^{r} & \sigma_{e2}^{r}
	\end{pmatrix}
	\begin{pmatrix}
	v_{e1}^{l} & v_{e2}^{l} \\
	\sigma_{e1}^{l} & \sigma_{e2}^{l}
	\end{pmatrix}^{-1},
	\end{equation}
or in its full, expanded form
	\begin{equation}
	\mathsf{T} = 
	\vect{\zeta}^{b} \vect{\delta}^{b}(x^r, x^b) 
	\begin{pmatrix}
	A_{e1}^{b+} & A_{e2}^{b+} \\
	A_{e1}^{b-} & A_{e2}^{b-}
	\end{pmatrix}
	\begin{pmatrix}
	A_{e1}^{a+} & A_{e2}^{a+} \\
	A_{e1}^{a-} & A_{e2}^{a-}
	\end{pmatrix}^{-1}
	(\vect{\delta}^{a}(x^l, x^a))^{-1} (\vect{\zeta}^{a})^{-1}.
	\end{equation}
The scattering matrix is calculated from the transfer matrix. 
Following \cite{amirkhizi_homogenization_2017} and for a reciprocal and symmetric system (i.e. $\mathsf{T}_{11} = \mathsf{T}_{22}$ and $\det \mathsf{T} = \mathsf{T}_{11} \mathsf{T}_{22} - \mathsf{T}_{12} \mathsf{T}_{21} = 1$) with de-embedding locations at interfaces $x^a = x^l$ and $x^b = x^r$:
	\begin{equation} \label{eq:delta}
	\Delta=\mathsf{T}_{21}+(\mathsf{T}_{11}+\mathsf{T}_{22})Z_{0}+\mathsf{T}_{12}Z_{0}^{2},
	\end{equation}
  \begin{equation}
  	\mathsf{S}^v_{ab} = \mathsf{S}^v_{ba} = \mathsf{S}^\sigma_{ab} = \mathsf{S}^\sigma_{ba} = \frac{2Z_{0}}{\Delta},
  \end{equation}
	\begin{equation}
    \mathsf{S}^v_{aa} = \mathsf{S}^v_{bb} = -\mathsf{S}^\sigma_{aa} = -\mathsf{S}^\sigma_{bb} = -\frac{\mathsf{T}_{21}-\mathsf{T}_{12}Z_{0}^{2}}{\Delta},
	\end{equation}
where superscripts $v$ or $\sigma$ are used when the wave amplitudes $A$ are associated with velocity or stress/pressure, respectively. The extraction of the remaining effective properties uses the notations: impedance ($Z$), wave number ($k$), wave phase velocity ($c_{\phi}$), density ($\rho$), and longitudinal modulus ($L$). 
The effective impedance and wave number may be extracted from the transfer matrix. With symmetry and reciprocity enforced as discussed earlier, the transfer matrix can be written as:
	\begin{equation}
	\mathsf{T} = 
	\begin{pmatrix}
	\cos kd   & \dfrac{i}{Z} \sin kd \\
	i Z \sin kd & \cos kd
	\end{pmatrix},
	\end{equation}
and therefore two equal and opposite solutions for impedance and wave number may be derived (note the reverse order of $\mp$ in the right hand side of Equation~\ref{eq:eik}):
  \begin{equation}
		Z^{\pm} = \frac{ \pm \sqrt{\mathsf{T}_{12}\mathsf{T}_{21} }}{\mathsf{T}_{12}},
		\label{eq:Z}
	\end{equation}
	\begin{equation}
	  e^{-ik^{\pm}d} = \mathsf{T}_{11} \mp \sqrt{\mathsf{T}_{12}\mathsf{T}_{21}},
	  \label{eq:eik}
	\end{equation}
	\begin{equation}
	k^{\pm}=\frac{\ln(e^{-ik^{\pm}d})}{-id},
	\label{wave_vector}
	\end{equation}
where $d$ is the sample thickness. Note that the real parts of $k^\pm$ have $2\pi n^\pm /d$ ambiguity ($n^\pm \in \mathbb{Z}$). The branch number ($n^\pm$) has been used to maintain the continuity of the wave number as a function of frequency (and by extension the apparent constitutive parameters when possible) in the past, especially when the structure is not fully elastic (lossless) \cite{abedi_use_2020}. For lossless systems, the use of only continuity will not remove the ambiguity. Instead, it is suggested that a lossy system with similar properties are modeled and then the value of loss is made to approach zero \cite{amirkhizi_homogenization_2017}. As this limiting process progresses, one must maintain the major features and shapes of the resulting functions. The limiting functions are physically expected to be the best representations of the lossless system. Further detail of the branch selection when the continuity method fails are discussed in Section~\ref{branch_selection} and \cite{abedi_use_2020}, but all of the results in Section~\ref{results_discussion} are derived using only continuity method. This formulation is derived for a symmetric and reciprocal specimen, i.e. $Z^- = -Z^+ = -Z$ and $k^- = -k^+ = -k$. The phase advance $Q$ through a single unit cell of thickness $a_x = d/N$, where $N$ is the number of repeating unit cells, is
	\begin{equation}
	Q = \frac{k d}{N},
	\label{Q_eq}
	\end{equation}
and the phase velocity as a function of frequency ($f$) can be calculated as
	\begin{equation} \label{eq:cp}
	c_\phi = \frac{2 \pi f}{k}.
	\end{equation}
The effective density and (longitudinal) modulus of the specimen are written as 
	\begin{equation}
  \rho = \frac{Z}{c_\phi},
	\end{equation}
	\begin{equation}
  L = c_\phi Z = \frac{Z^2}{\rho}.
	\label{dynamic_L}
	\end{equation}
If the specimen is asymmetric or non-reciprocal, overall coupling parameters will be required and details of such analysis may be found in \cite{amirkhizi_homogenization_2017}.

\section{Results and Discussion} \label{results_discussion}

\begin{figure}[!ht]
		\centering
		\includegraphics [width=\textwidth]{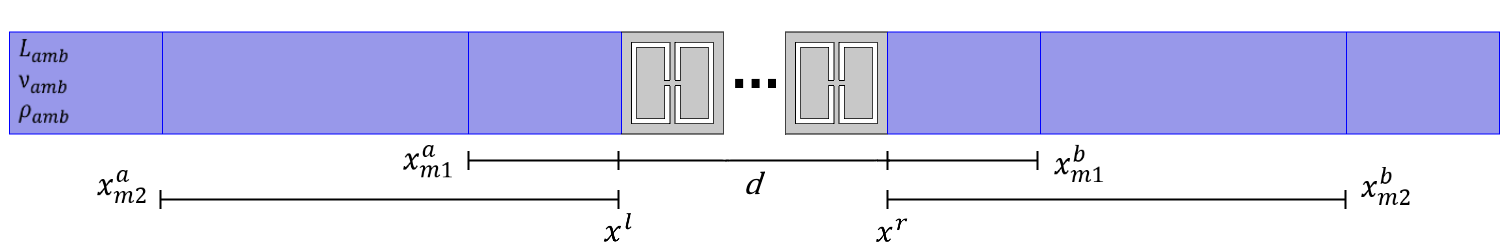}
		\caption{The model is constructed in COMSOL Multiphysics with the ambient media ($L_{amb}$, $\nu_{amb}$, and $\rho_{amb}$) surrounding the metamaterial ($d = N a_x$).}
		\label{comsol} 
	\end{figure}

Two dimensional (plane strain) COMSOL frequency domain simulations of the specimens placed between two ambient homogeneous media of finite length, as shown in Figure~\ref{comsol}, were performed over the range of \SIrange{0.5}{5}{kHz}. The top and bottom boundaries are considered periodic, while the left and right boundaries have harmonic excitations. Further detail of this modeling can be found in \cite{aghighi_low-frequency_2019}. The numerical mesh convergence was assured by observing models with maximum mesh element sizes of 0.5, 1, and 2 \si{mm}. As a verification of the analysis approach, the results calculated using numerical integration of velocity and those of traction at the measurement points were compared and proven to be identical. The proper adjustment (Equation~\ref{eq:Z_vel} or \ref{eq:Z_stress}) was made for the corresponding measurement type. In the following, the effect of ambient domain material properties and number of cells across the thickness of the slab, $N$, are studied. As $N$ increases, the apparent effective properties change but eventually converge. For thin slabs (e.g. $N = 1$), the effect of the ambient domain elastic stiffness, $L_{amb}$, is observed. However, if $L_{amb}$ is kept constant, changing the density of the ambient domain, $\rho_{amb}$, (or equivalently its wave speed or impedance) has no effect on the results.
  
\subsection{Effect of ambient parameters} \label{effect_ambient}

\begin{figure}[!ht]
		\centering
		\begin{subfigure}{0.49\linewidth} \centering
			\includegraphics[scale=0.15]{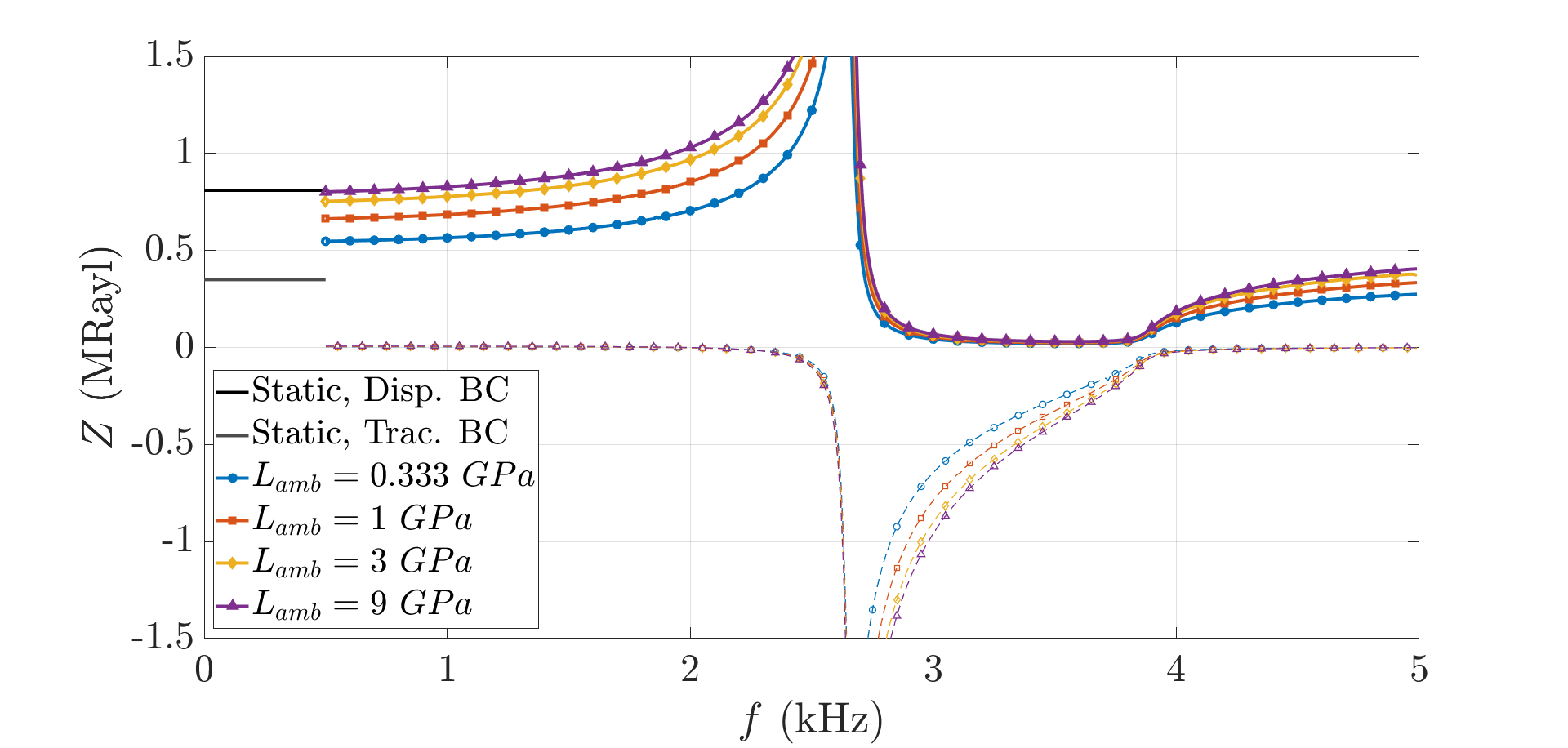}
			\caption{ }
			\label{Hmat_Z}
		\end{subfigure}
		\begin{subfigure}{0.49\linewidth} \centering
			\includegraphics [scale=0.15]{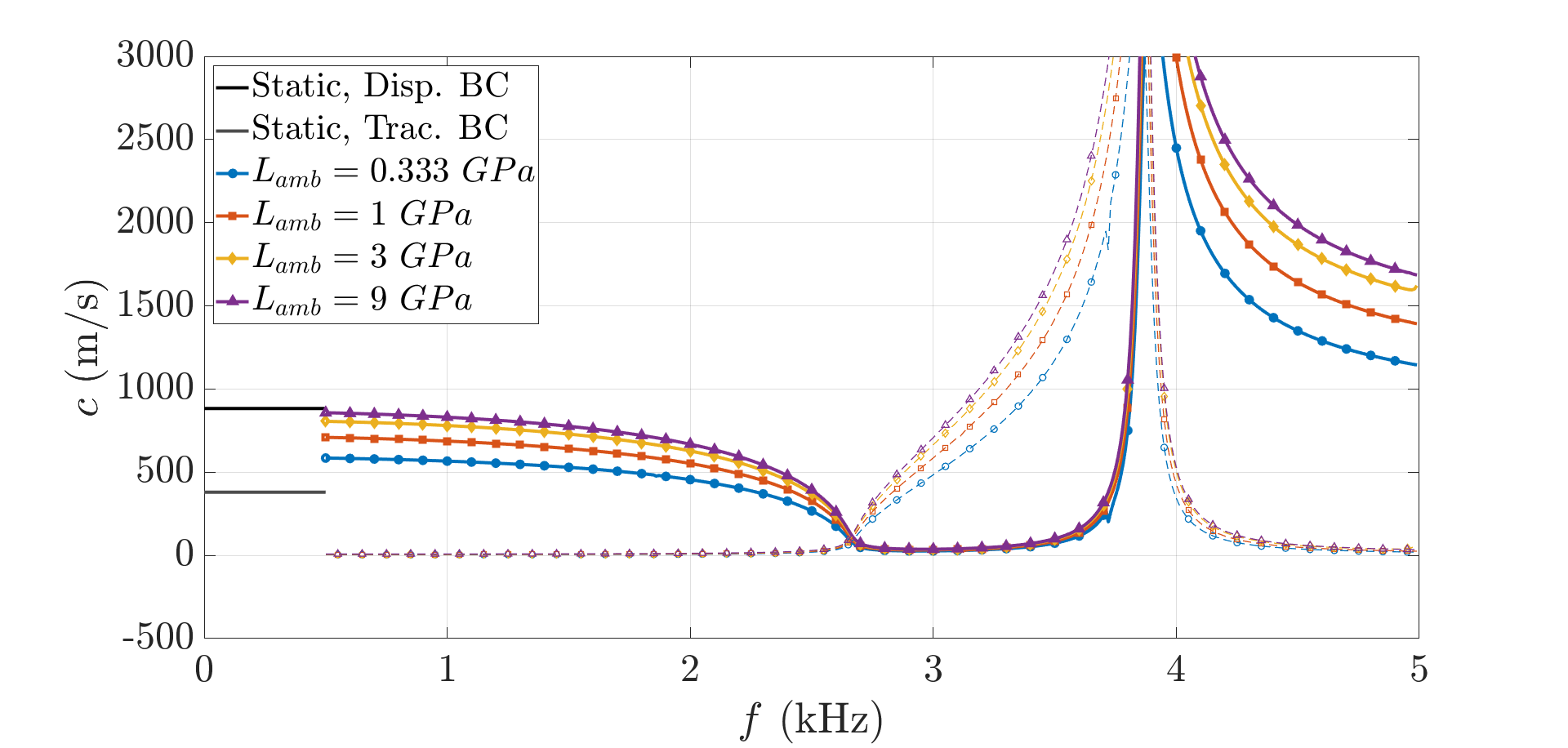}
			\caption{ }
			\label{Hmat_c}
		\end{subfigure}
		\begin{subfigure}{0.49\linewidth} \centering
			\includegraphics[scale=0.15]{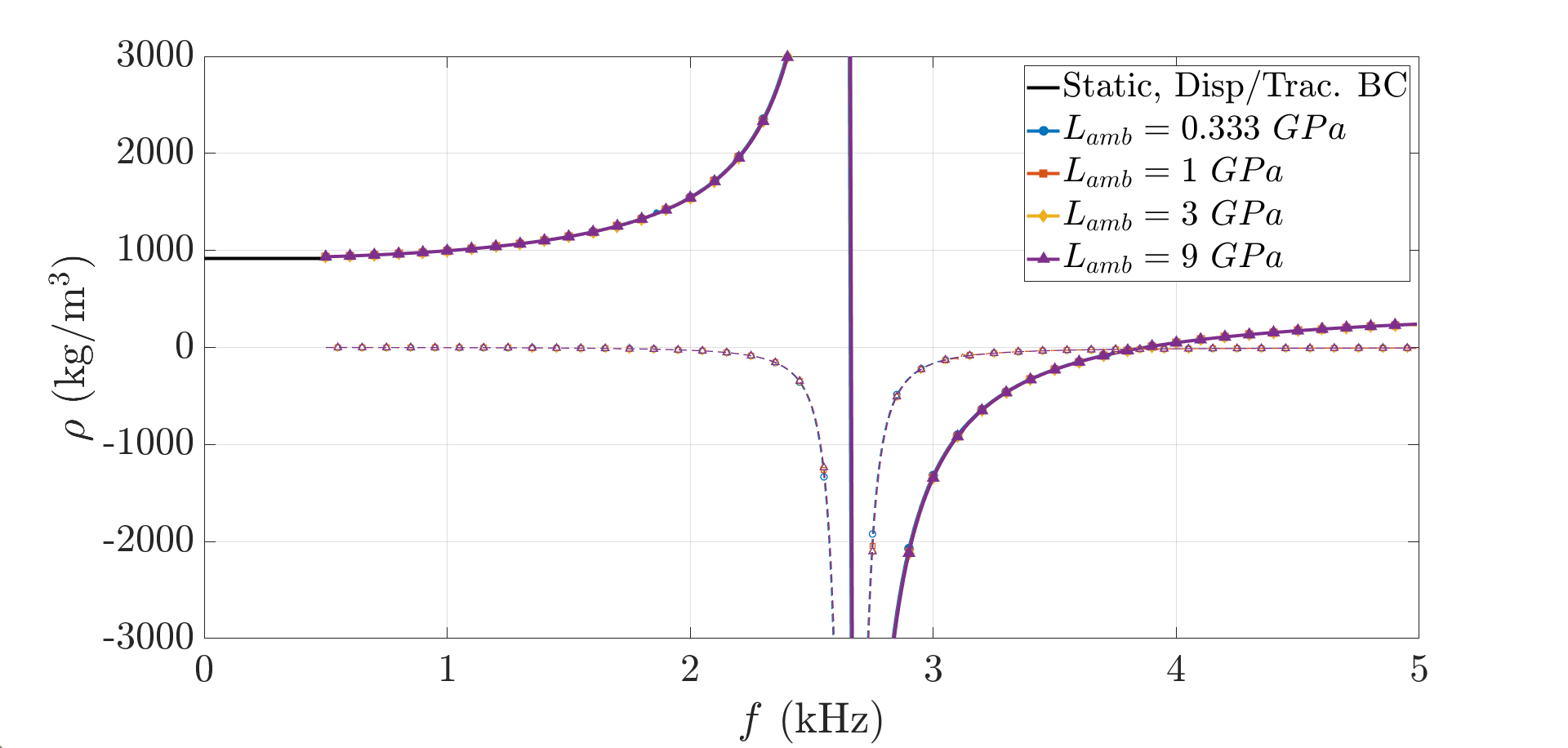}
			\caption{ }
			\label{Hmat_rho}
		\end{subfigure}
		\begin{subfigure}{0.49\linewidth} \centering
			\includegraphics [scale=0.15]{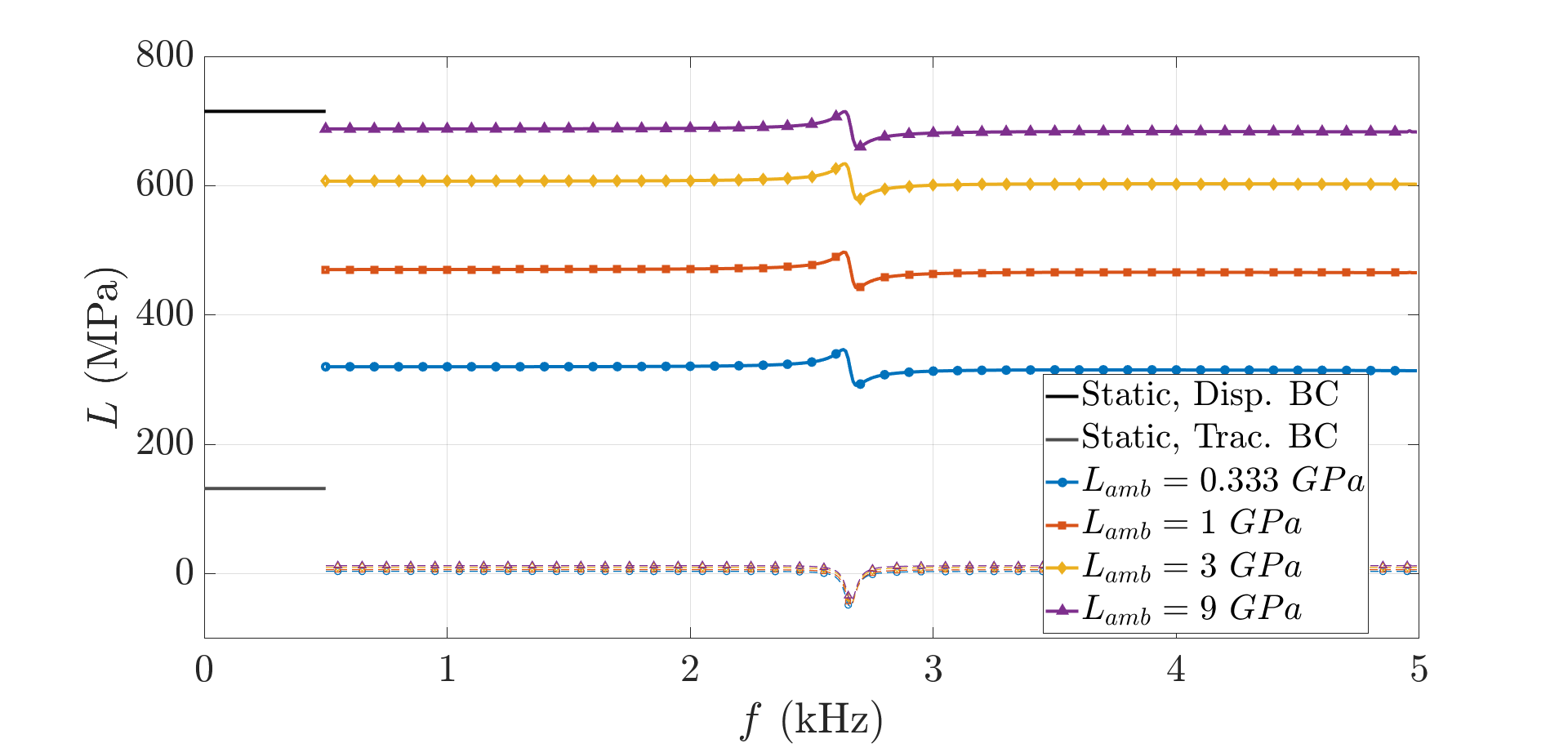}
			\caption{ }
			\label{Hmat_L}
		\end{subfigure}
		\caption{Effective properties $Z$, $c$, $\rho$, and $L$ of a finite array with $N = 1$ H Design unit cell for $L_{amb}=$ \SIlist{0.333;1;3;9}{GPa}. The line markers at 0 kHz identify the quasi-static bounds from Table~\ref{table_moduli}. The Normal Disp. BC also matches with the long wavelength limit approximation based on eigenfrequency analysis. The imaginary parts are included as thinner, dashed lines for each result.}
		\label{Hmat-properties} 
	\end{figure}

A number of the model's physical parameters were varied to determine their influence on the metamaterial slab's response and overall (apparent) constitutive parameters. The distance from the metamaterial surface and the measurement locations was changed with $x^{a}_{m1}=x^{b}_{m1}=1,2,5,10,20$ \si{cm} and $x^{a}_{m2}=x^{b}_{m2}=2,4,10,20,40$ \si{cm} with no observable effect on the result, indicating near-field effects do not appear significant. The known material properties of the ambient media were modified next. For a single-cell thick specimen, the longitudinal modulus and density of the ambient media were varied while their wave speed and impedance changed indirectly as functions of those set values (Equations~\ref{eq:speed} and \ref{eq:imp}). The longitudinal modulus was set to $L_{amb}=$ \SIlist{0.333;1;3;9}{GPa} combined with densities of $\rho_{amb} = $\SIlist{333;1000;3000}{kg/m^{3}}. This provides values both above and below the metamaterial's constituent resin properties of $L_{res} = $\SI{4.4}{GPa} and $\rho_{res} = $\SI{1161}{kg/m^{3}}. The Poisson's ratio in all cases was kept fixed $\nu_{amb} = 0.25$. It was determined that for the same ambient longitudinal modulus, variation of ambient density (or equivalently, wave speed, or impedance) does not change the calculated overall properties. However, when the longitudinal modulus of the ambient media is changed (with fixed density), the apparent metamaterial parameters are affected, as depicted in Figure~\ref{Hmat-properties}. The apparent behavior at the long wavelength limit with stiffer ambient media is closer to the eigenfrequency solution and static displacement BC model, while more compliant ambient media lead to long wavelength limit response closer to the estimates from static traction BC model. However, in all cases the location of the stop band (identified as the regions where extracted wave speed and impedance become imaginary-dominant) is identical to that of the eigenfrequency analysis. It must be emphasized that the extracted dispersive density, here and in all following cases, does not seem to be influenced by the change of the ambient media modulus. 
  
\subsection{Effect of number of cells through thickness} \label{effect_cell_number}
	
\begin{figure}[!ht]
		\centering
		\begin{subfigure}{0.49\linewidth} \centering
			\includegraphics[scale=0.15]{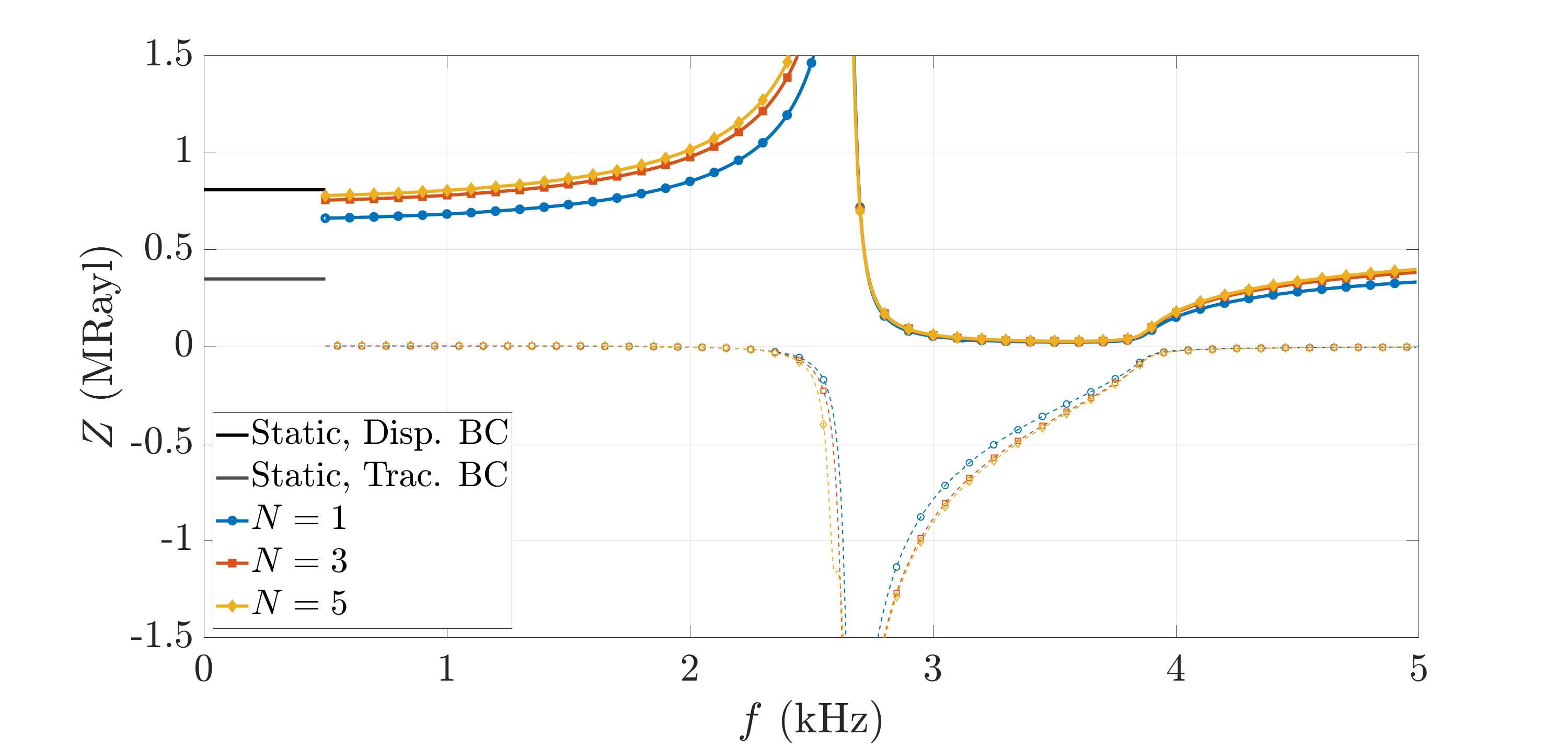}
			\caption{ }
			\label{Ncells_Z}
		\end{subfigure}
		\begin{subfigure}{0.49\linewidth} \centering
			\includegraphics [scale=0.15]{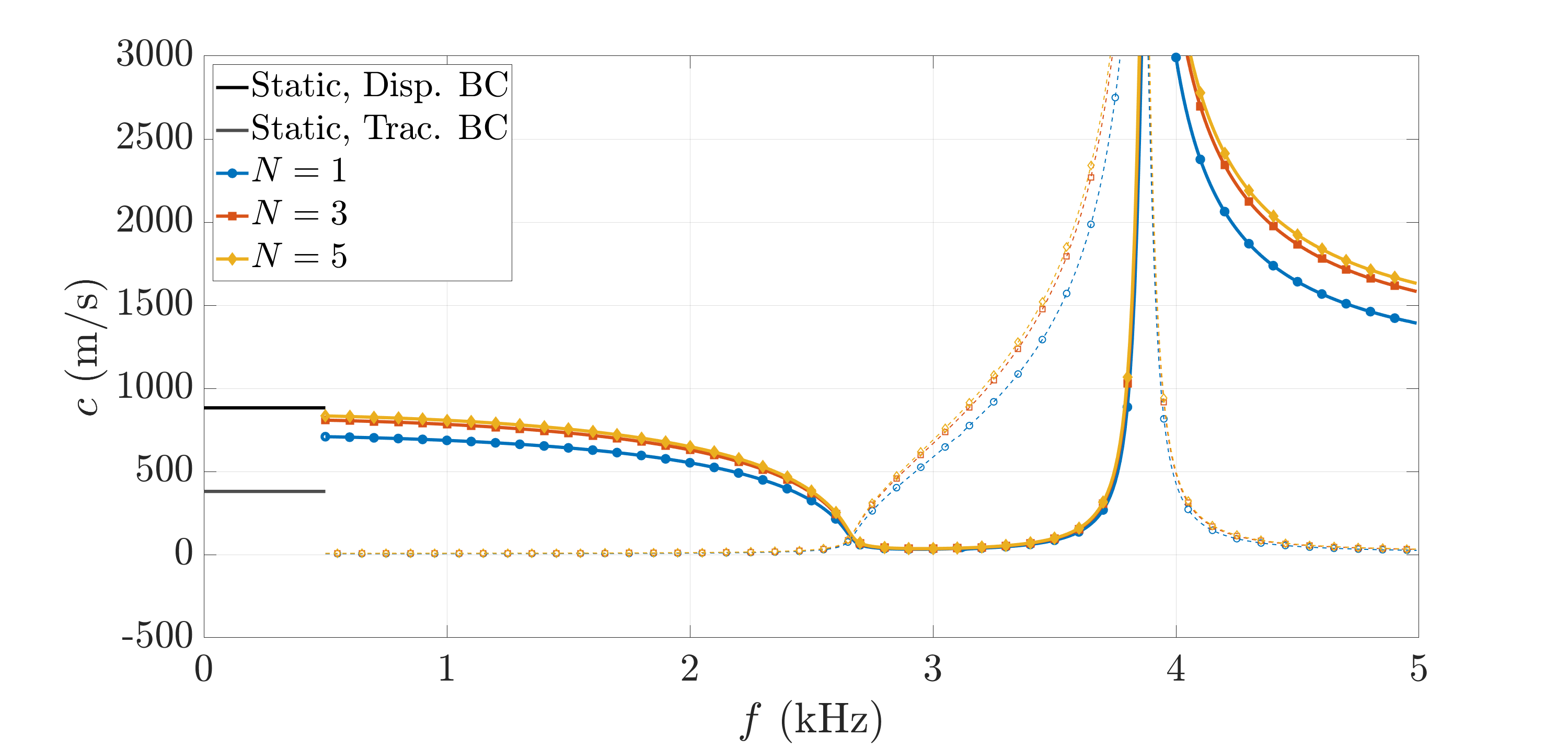}
			\caption{ }
			\label{Ncells_c}
		\end{subfigure}
		\begin{subfigure}{0.49\linewidth} \centering
			\includegraphics[scale=0.15]{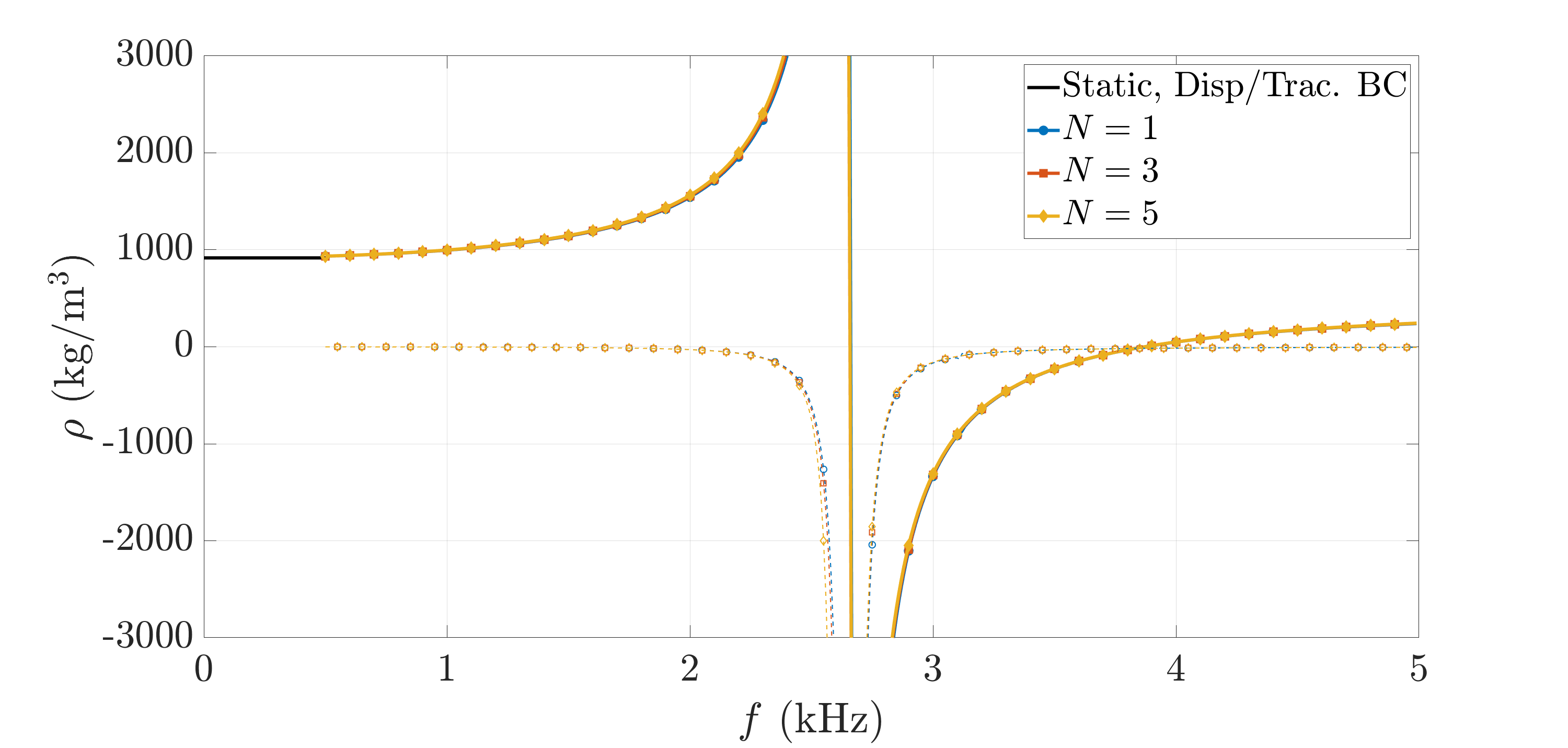}
			\caption{ }
			\label{Ncells_rho}
		\end{subfigure}
		\begin{subfigure}{0.49\linewidth} \centering
			\includegraphics [scale=0.15]{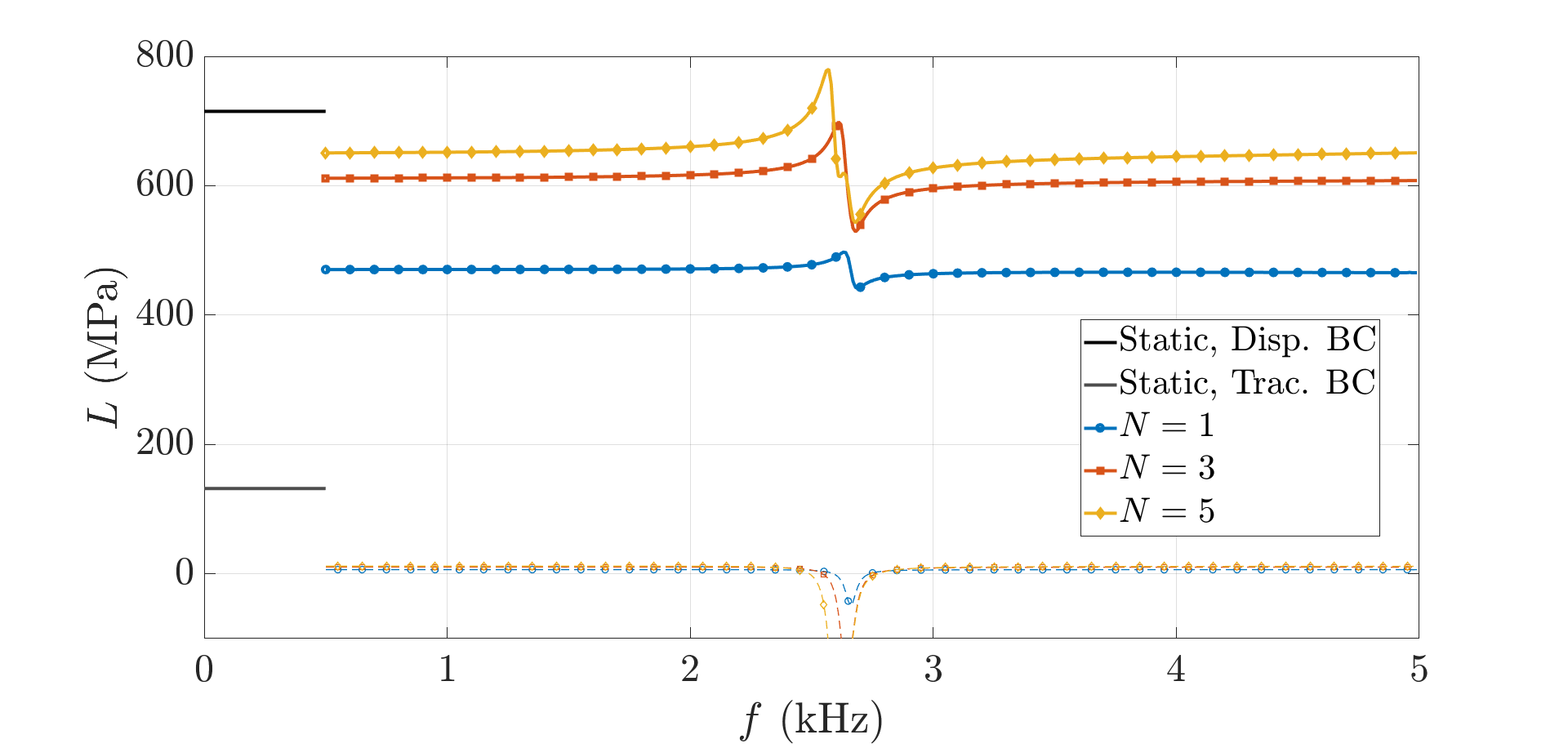}
			\caption{ }
			\label{Ncells_L}
		\end{subfigure}
		\caption{Extracted apparent impedance, wave speed, density, and longitudinal modulus for finite arrays with \numlist{1;3;5} of the \SI{10}{mm} H Design unit cells, with the assumption that interior and boundary regions are exactly the same. The imaginary parts are included as thinner, dashed lines for each result.}
		\label{HN-properties} 
	\end{figure}

Intuitively, the thickness of the metamaterial specimen (or number of repeating unit cells in thickness direction) should play a significant role in performance, though ideally the effective properties should be independent of geometry. The number of repeating cells $N$ was varied among 1, 3, and 5, with the corresponding solutions presented in Figure~\ref{HN-properties}. In this case $L_{amb}=1$ GPa and $\rho_{amb}=1000$ kg/m$^3$.  Unsurprisingly, solutions with more cells converge towards the eigenfrequency (representing infinitely periodic system) and static displacement BC solutions in the long wavelength limit. The behavior within the stop band is largely unaffected by the number of unit cells used in the model (except for apparent modulus), while the extracted properties outside of the band are sensitive to this variation. While the extracted values do appear to converge quickly with increasing $N$, the skin (boundary) cells are influenced by the ambient domain and in the following we seek to assign different apparent properties to them compared to interior cells. This approach is shown to resolve the dependence on ambient media modulus satisfactorily as well, as the interior region properties have effectively no dependence on ambient media while the skin region properties naturally do.
	
\section{Distinct Effective Properties at the Skin and Interior Domains} \label{boundary_cells}

The assumption of uniform properties throughout the slab in the previous section led to the apparent properties that depend on the modulus of the ambient media or the number of cells across the slab. The increase in the modulus of ambient media or the increase in the number of cells in the array lead to convergence of these apparent constitutive properties towards the eigenfrequency estimates, which are essentially associated with infinite arrays. As the number of cells through the thickness increases, a finite array's behavior becomes dominated by the interior cells and more closely matches the boundary conditions of the infinitely periodic eigenfrequency model. The increase of the ambient media modulus leads to similar effect. Combination of ambient influence in Figure~\ref{Hmat-properties} with thickness-dependency in Figure~\ref{HN-properties} prompts consideration of different material properties for interior and skin (boundary cells) regions. In this study, the skin region is considered to be equivalent to a single cell. The skin region properties may be considered a function of both the cell micro-structure as well as the ambient domain, while the interior region's extracted properties should ideally be independent of the ambient domain constitutive parameters. Furthermore, it is desirable that neither the properties of the skin region, nor those of the interior domain, would depend on the total thickness of the slab. In other words, the scattering off any slab thickness is reproduced based on these sets of constitutive properties. The process starts with considering different transfer matrices for the two domains. Assuming for simplicity that the ambient media on either side are identical, the transfer matrix of the left and right skin regions, although each potentially asymmetric, may be related to each other through parity conditions \cite{wang_exceptional_2022}. The transfer matrix for a two cell array may be written in terms of that of the skin region only as
\begin{equation}
	\mathsf{T}^{N=2} = \mathsf{T}^{out} F (\mathsf{T}^{out})^{-1} F, 
	\label{eq:2H}
\end{equation}
where $F$ is the parity operator 
\begin{equation}
	F = \begin{pmatrix} -1   & 0 \\	0 & 1 \end{pmatrix}, 
	\label{eq:parity}
\end{equation}
and $\mathsf{T}_{out}$ is the transfer matrix skin (boundary) cell at the right end of the array. 
$\mathsf{T}_{N=2}$ may be directly measured or evaluated from numerical simulations and can then be used in Eq.~\ref{eq:2H} to determine the skin transfer matrix. 
The equations for impedance, $Z_{out}$, and normalized phase advance through a single cell, $k_{out}a_x$, for the outer skin cells reduce nicely as
\begin{equation} \label{eq:kaout}
    k_{out}a_x = \frac{\log \left(\pm\sqrt{\mathsf{T}_{11}^{N=2}+w_0}\right)}{i},
\end{equation}
\begin{equation} \label{eq:Zout}
    Z_{out}=\frac{w_0}{\mathsf{T}_{12}^{N=2}},
\end{equation}
where
\begin{equation} \label{eq:x0}
    w_0=\pm\sqrt{-1+(\mathsf{T}_{11}^{N=2})^2}.
\end{equation}
It must be noted that these forms do not represent the only possible solution, and they are derived with the assumption that the properties of the skin region are symmetric, in particular the impedance. Without this assumption, there exists an infinite class of potential solutions that can be parameterized by a single parameter, e.g. the phase advance $k_{out}a_x$ which unlike the effective impedance is required to be the same in both directions and sides due to the reciprocity. Additionally, within this form exists a series of four potential values, denoted by the $\pm$ signs in Eq.~\ref{eq:x0} and \ref{eq:kaout}. The sign option for the square root in $w_0$ must be kept consistent for corresponding $k_{out}a_x$ and $Z_{out}$ solutions. To decide which of the four values is to be used, the continuity is enforced starting at $f\rightarrow 0$.

The transfer matrix for a single inner cell can then be extracted from the three cell array result based on 
 \begin{equation}
	\mathsf{T}^{N=3} = \mathsf{T}^{out} \mathsf{T}^{in} F (\mathsf{T}^{out})^{-1} F.
	\label{eq:3H}
\end{equation}
The solution for the interior domain has a closed form based on the two and three cell transfer matrices, though it looks more complex and is broken apart as follows
\begin{equation} \label{eq:x1}
\begin{aligned}
    w_1={} & (\mathsf{T}_{12}^{N=2})^2-(\mathsf{T}_{11}^{N=3})^2\mathsf{T}_{12}^{N=2} \\
    & + 2 \mathsf{T}_{11}^{N=2} \mathsf{T}_{12}^{N=2} \mathsf{T}_{11}^{N=3} \mathsf{T}_{12}^{N=3} \\
    & + (\mathsf{T}_{12}^{N=3})^2-(\mathsf{T}_{11}^{N=2})^2 (\mathsf{T}_{12}^{N=3})^2,
\end{aligned}
\end{equation}
\begin{equation} \label{eq:x2}
    w_2=\pm\sqrt{-4 (\mathsf{T}_{12}^{N=2})^2 (\mathsf{T}_{12}^{N=3})^2 + w_1^2}, 
\end{equation}
\begin{equation} \label{eq:x3}
\begin{aligned}
    w_3={} & -4 \mathsf{T}_{12}^{N=2}((-1+(\mathsf{T}_{11}^{N=3})^2)(\mathsf{T}_{12}^{N=2})^2 \\
    & -2 (1+\mathsf{T}_{11}^{N=2})\mathsf{T}_{11}^{N=3}\mathsf{T}_{12}^{N=2}\mathsf{T}_{12}^{N=3}  \\
    & +(1+\mathsf{T}_{11}^{N=2})^2 (\mathsf{T}_{12}^{N=3})^2), 
\end{aligned}
\end{equation}
\begin{equation} \label{eq:eps}
    e^{ik_{in}a_x}=\frac{w_1 + w_2}{2\mathsf{T}_{12}^{N=2}\mathsf{T}_{12}^{N=3}},
\end{equation}
\begin{equation} \label{eq:kain}
    k_{in}a_x = \frac{\log e^{ik_{in}a_x}}{i},
\end{equation}
\begin{equation} \label{eq:Zin}
    Z_{in}=\frac{(1+\mathsf{T}_{11}^{N=2}) w_2}{w_3}.
\end{equation}
The result for impedance is in fact associated with the particular symmetric assumption for the skin regions. However, the phase advance solution for the interior domain happens to be independent of that assumption. In other words, there is a a unique value of $k_{in}$ that can be determined based on the 2 and 3 cells scattering measurements or simulations. This derivation too requires a sign choice, presented by the $\pm$ in $w_2$. Continuity constraints should be enforced in selection of one solution and phase unwrapping as discussed in more detail in the next section.

\begin{figure}[!ht]
		\centering
		\begin{subfigure}{0.49\linewidth} \centering
			\includegraphics[scale=0.15]{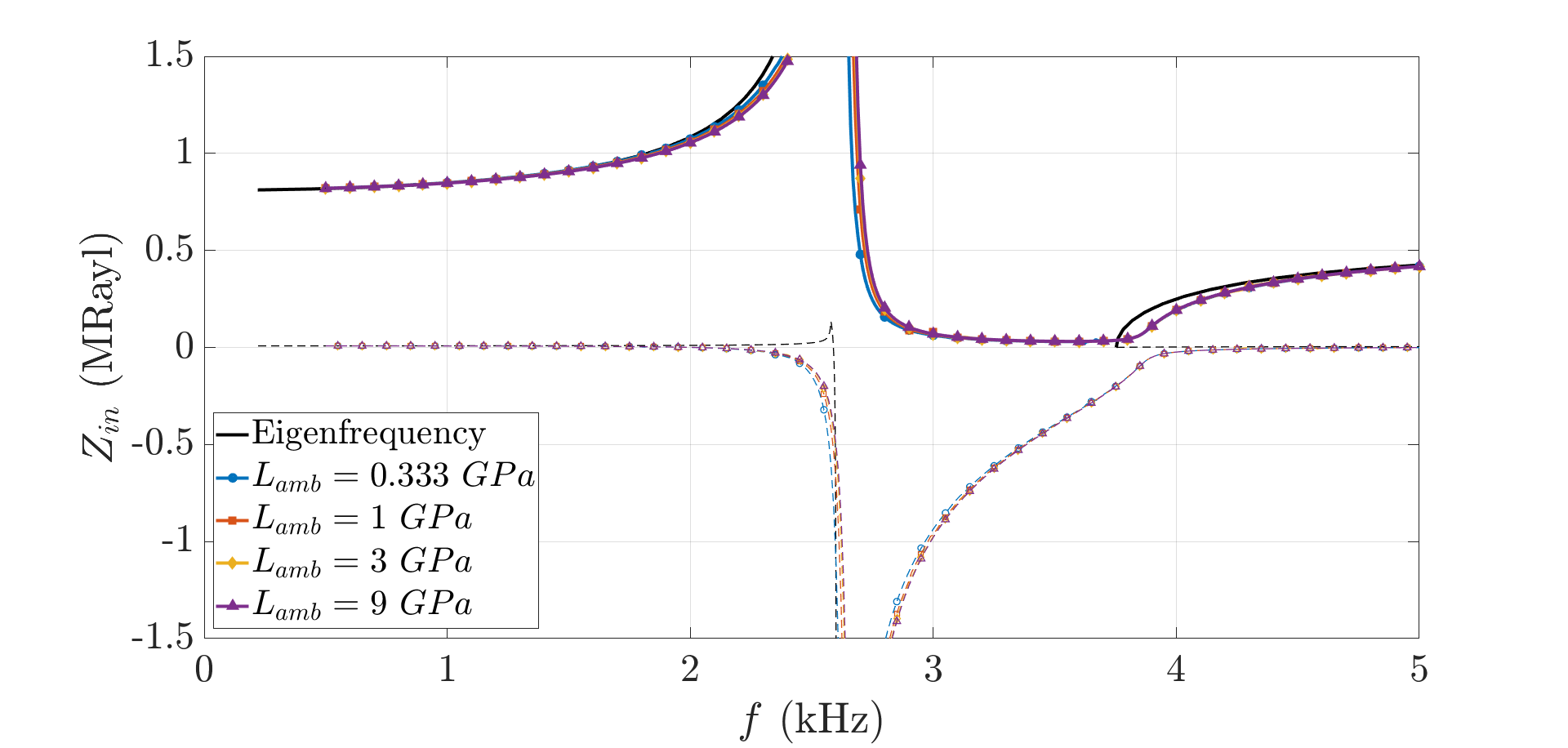}
			\caption{ }
			\label{Z_in}
		\end{subfigure}
		\begin{subfigure}{0.49\linewidth} \centering
			\includegraphics [scale=0.15]{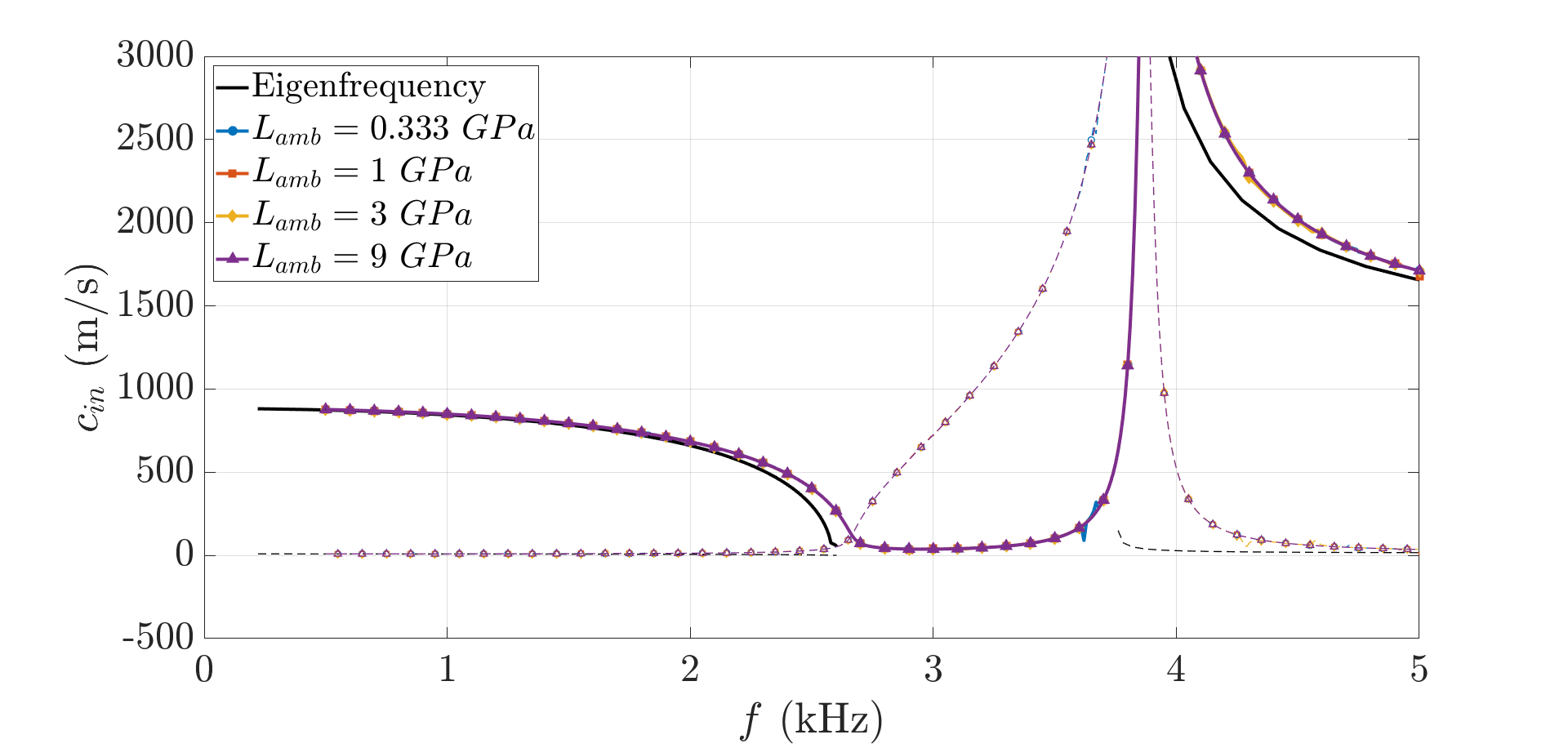}
			\caption{ }
			\label{c_in}
		\end{subfigure}
		\begin{subfigure}{0.49\linewidth} \centering
			\includegraphics[scale=0.15]{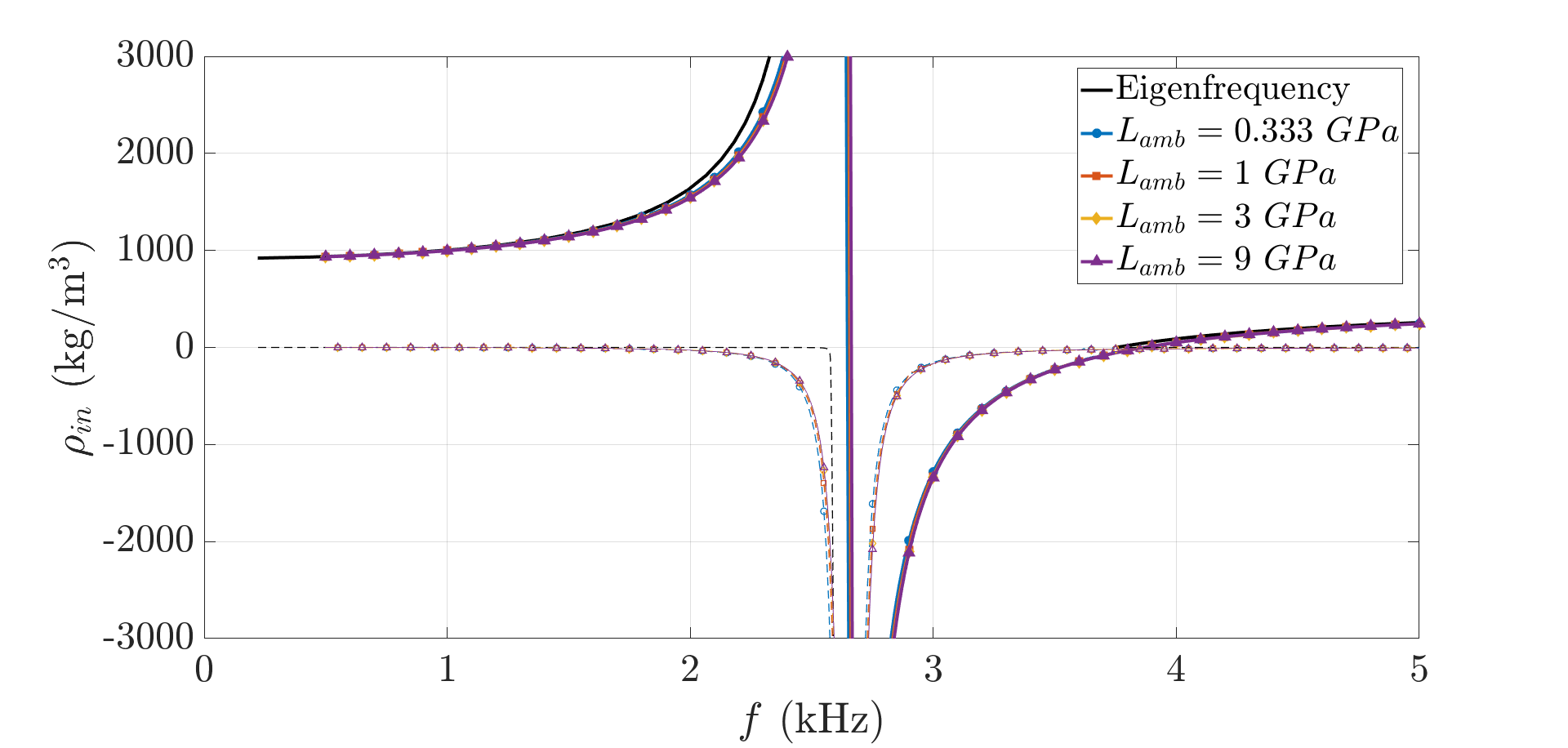}
			\caption{ }
			\label{rho_in}
		\end{subfigure}
		\begin{subfigure}{0.49\linewidth} \centering
			\includegraphics [scale=0.15]{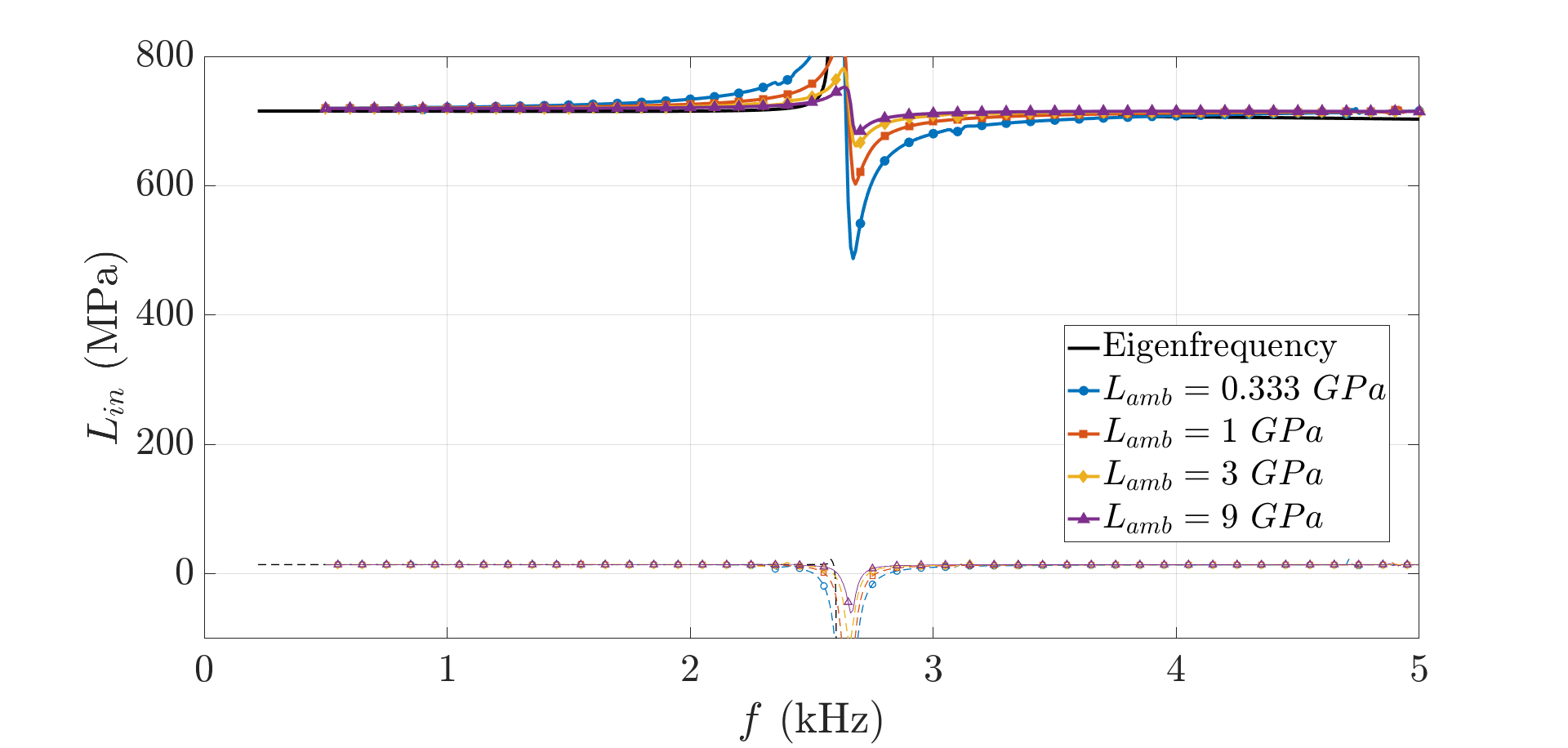}
			\caption{ }
			\label{L_in}
		\end{subfigure}
		\caption{Interior domain effective properties (\subref{Z_in}) impedance $Z_{in}$, (\subref{c_in}) phase velocity $c_{in}$,  (\subref{rho_in}) density $\rho_{in}$, and (\subref{L_in}) modulus $L_{in}$ are independent from ambient media (and number of cells). All $L_{amb}$ values produce overlapping curves. The imaginary parts are included as thinner, dashed lines for each result.}
		\label{Zc_in} 
	\end{figure}

\begin{figure}[!ht]
		\centering
		\begin{subfigure}{0.49\linewidth} \centering
			\includegraphics[scale=0.15]{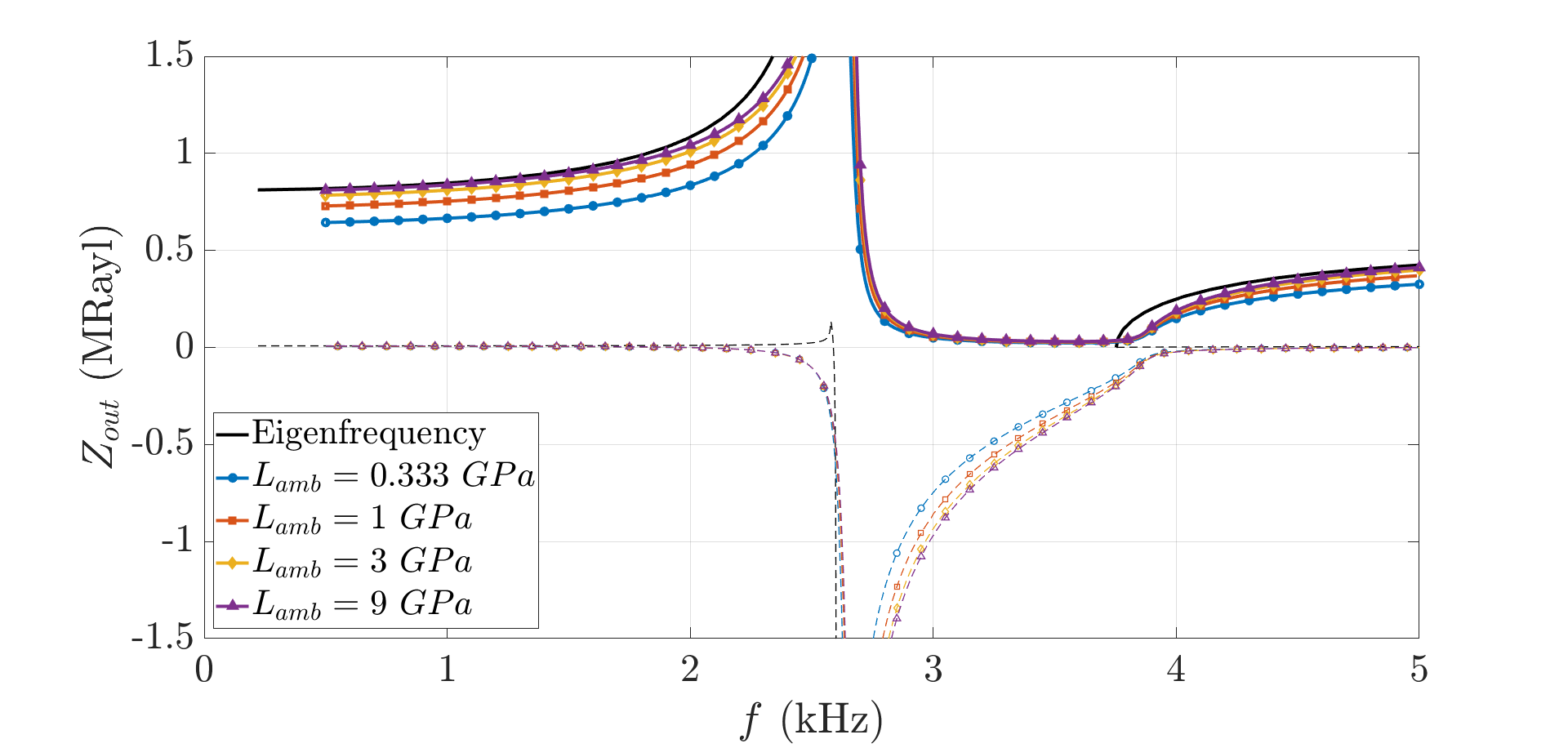}
			\caption{ }
			\label{Z_out}
		\end{subfigure}
		\begin{subfigure}{0.49\linewidth} \centering
			\includegraphics [scale=0.15]{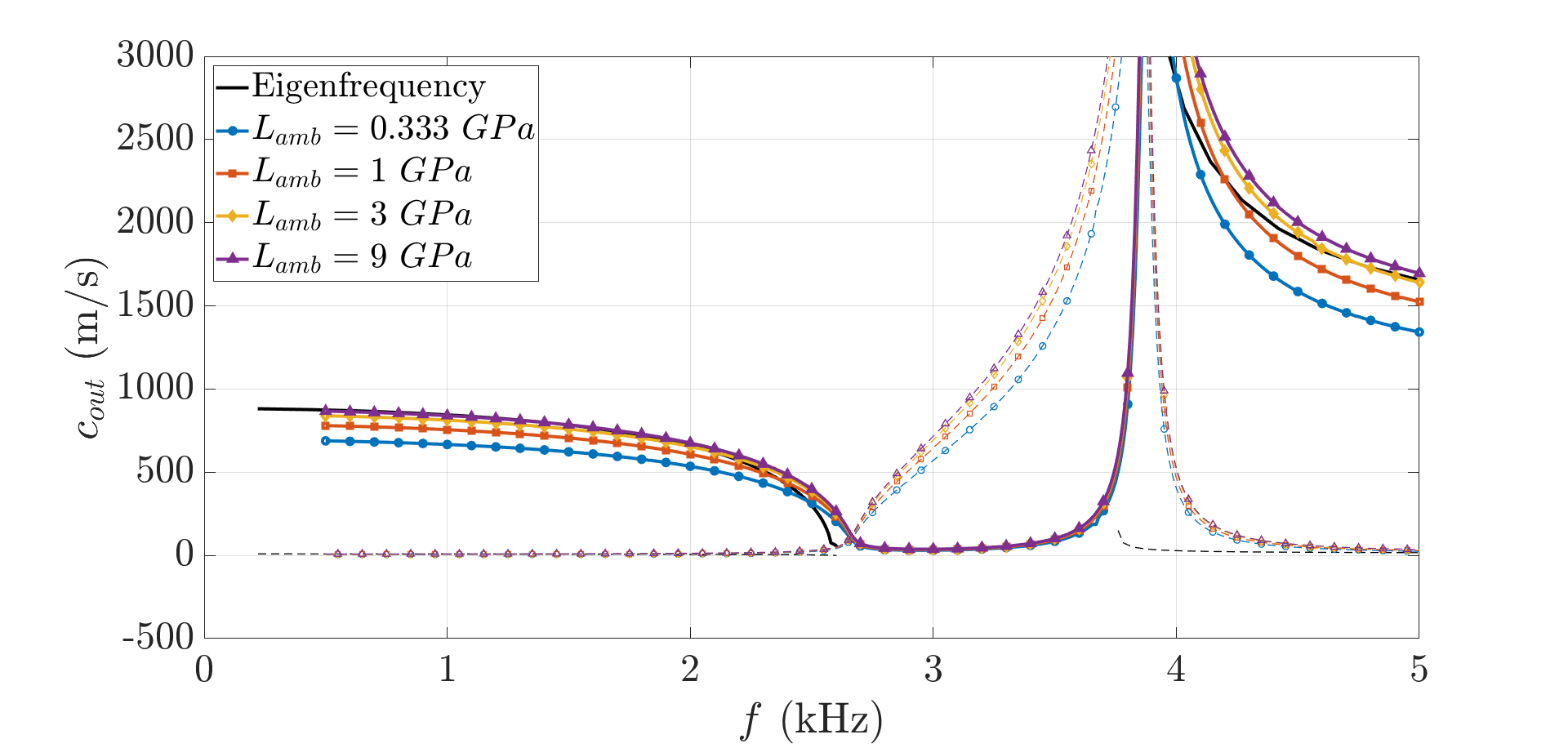}
			\caption{ }
			\label{c_out}
		\end{subfigure}
		\begin{subfigure}{0.49\linewidth} \centering
			\includegraphics[scale=0.15]{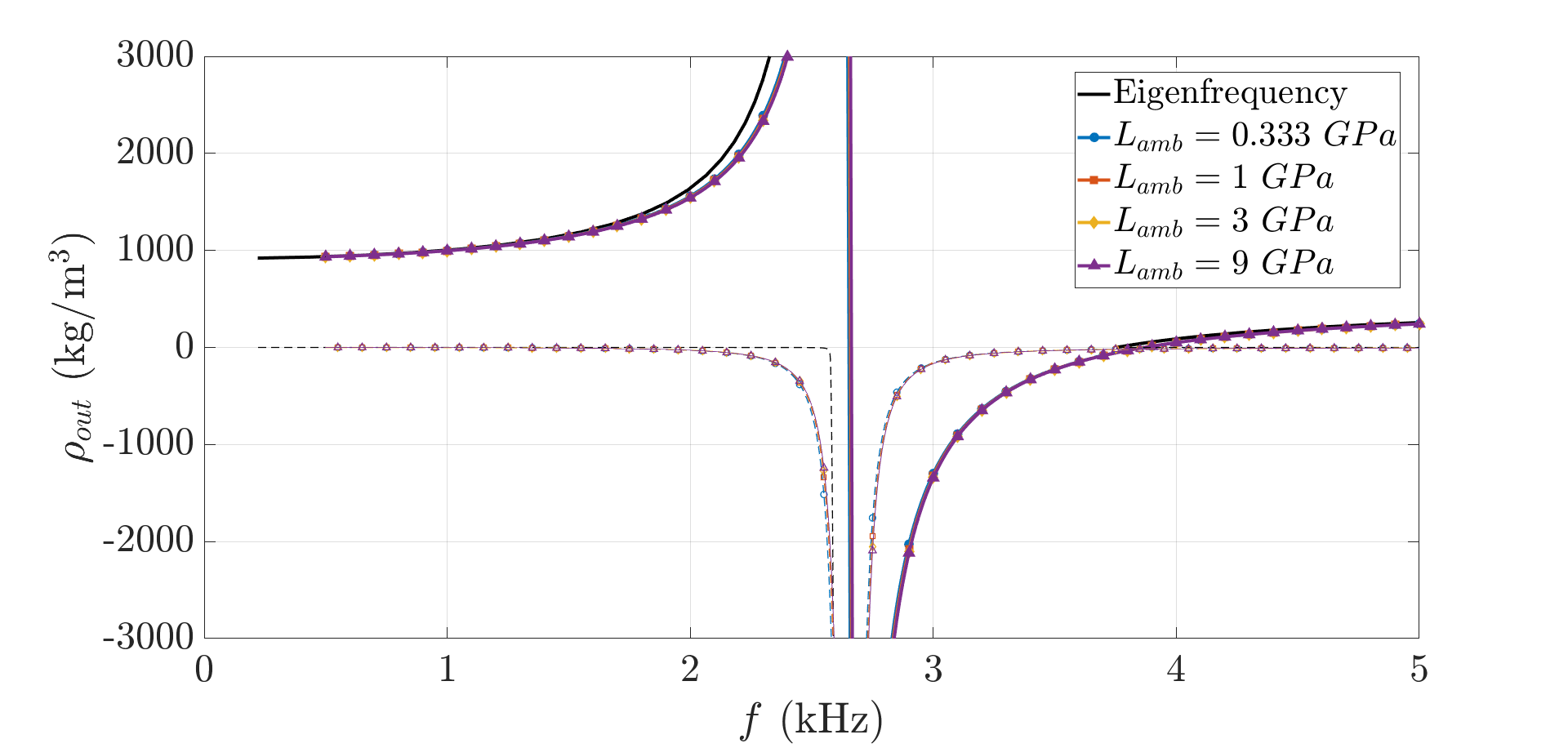}
			\caption{ }
			\label{rho_out}
		\end{subfigure}
		\begin{subfigure}{0.49\linewidth} \centering
			\includegraphics [scale=0.15]{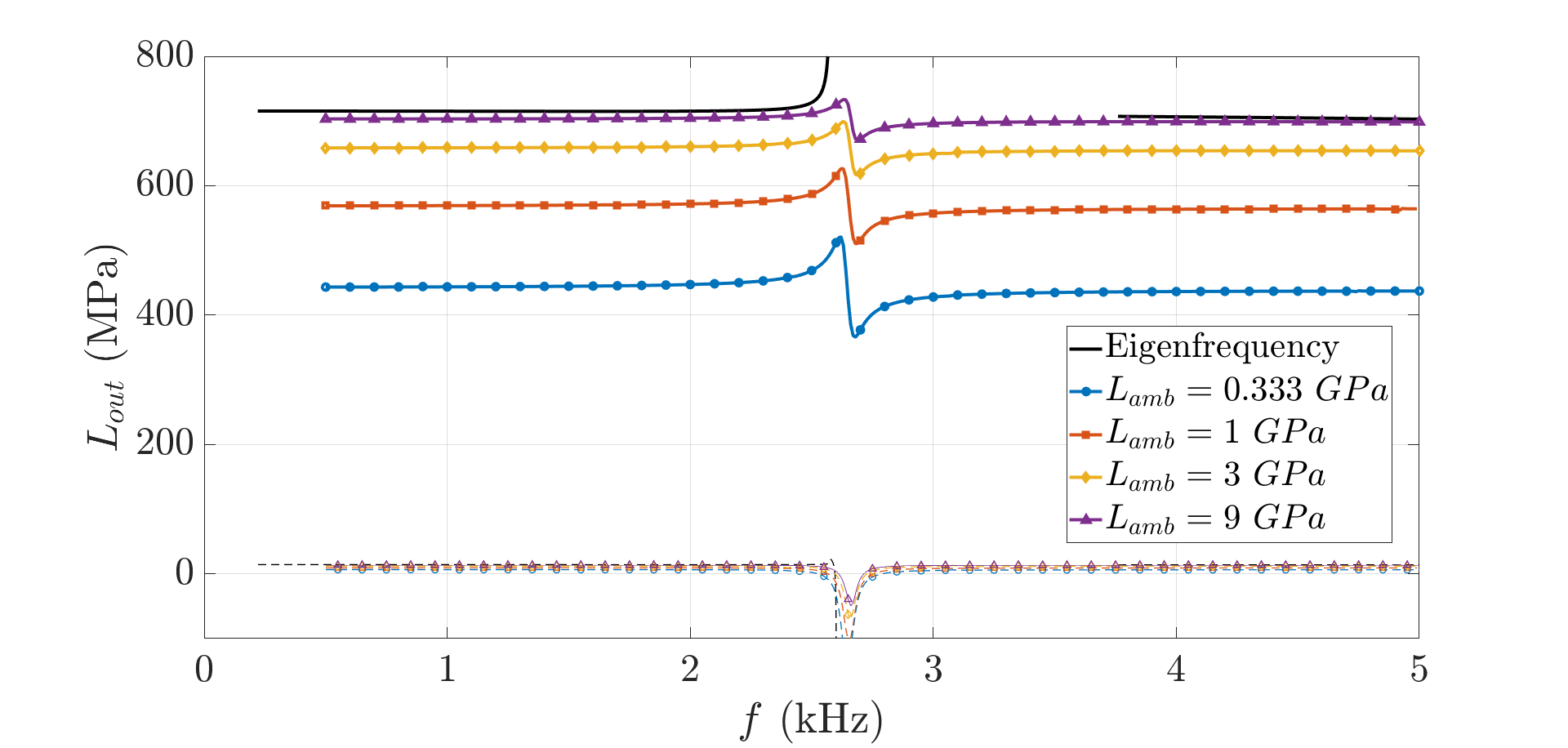}
			\caption{ }
			\label{L_out}
		\end{subfigure}
		\caption{Boundary cell properties (\subref{Z_out}) impedance $Z_{out}$ and (\subref{c_out}) phase velocity $c_{out}$ are functions of the ambient media modulus. When used to calculate (\subref{rho_out}) density $\rho_{out}$, it becomes independent, whereas, (\subref{L_out}) modulus $L_{out}$ experiences compounded effects. The imaginary parts are included as thinner, dashed lines for each result.}
		\label{Zc_out} 
	\end{figure}

The distinct $\mathsf{T}^{out}$ and $\mathsf{T}^{in}$ transfer matrices are confirmed to produce the scattering response of arrays with higher number of cells $N>3$, with very good accuracy (not shown here). The values associated with the interior region as calculated with Eq.~\ref{eq:kain} and \ref{eq:Zin} remain largely independent of the ambient media properties as seen in Figures~\ref{Z_in} and \ref{c_in} and match very closely with eignefrequency results outside of the stop band where such values are available. Note that the impedance from the eigenfrequency simulations is calculated based on line integration of normal traction and particle velocity components on the cells vertical faces:
\begin{equation}
    Z_{eig}=-\frac{\int \sigma_{xx} \,dy}{\int v_x \,dy},
\end{equation}
where integrals are performed at either $x$-face of the unit cells. Figures~\ref{Z_out} and \ref{c_out} show that the constitutive properties of skin or boundary regions, $Z_{out}$ and $c_{out}$, increase with $L_{amb}$. Again, the ambient media density was proven to have no influence over the response (not pictured). This is consistent with observations from Figures~\ref{Hmat-properties} and \ref{HN-properties}. Calculating effective densities and moduli based on these values (Figures~\ref{rho_out} and \ref{L_out}) shows that the effective density is still independent of the ambient medium properties as well, consistent with previous observations. 
	
\begin{figure}[!ht]
	\centering
		\begin{subfigure}{0.48\linewidth}
			\includegraphics[scale=0.15]{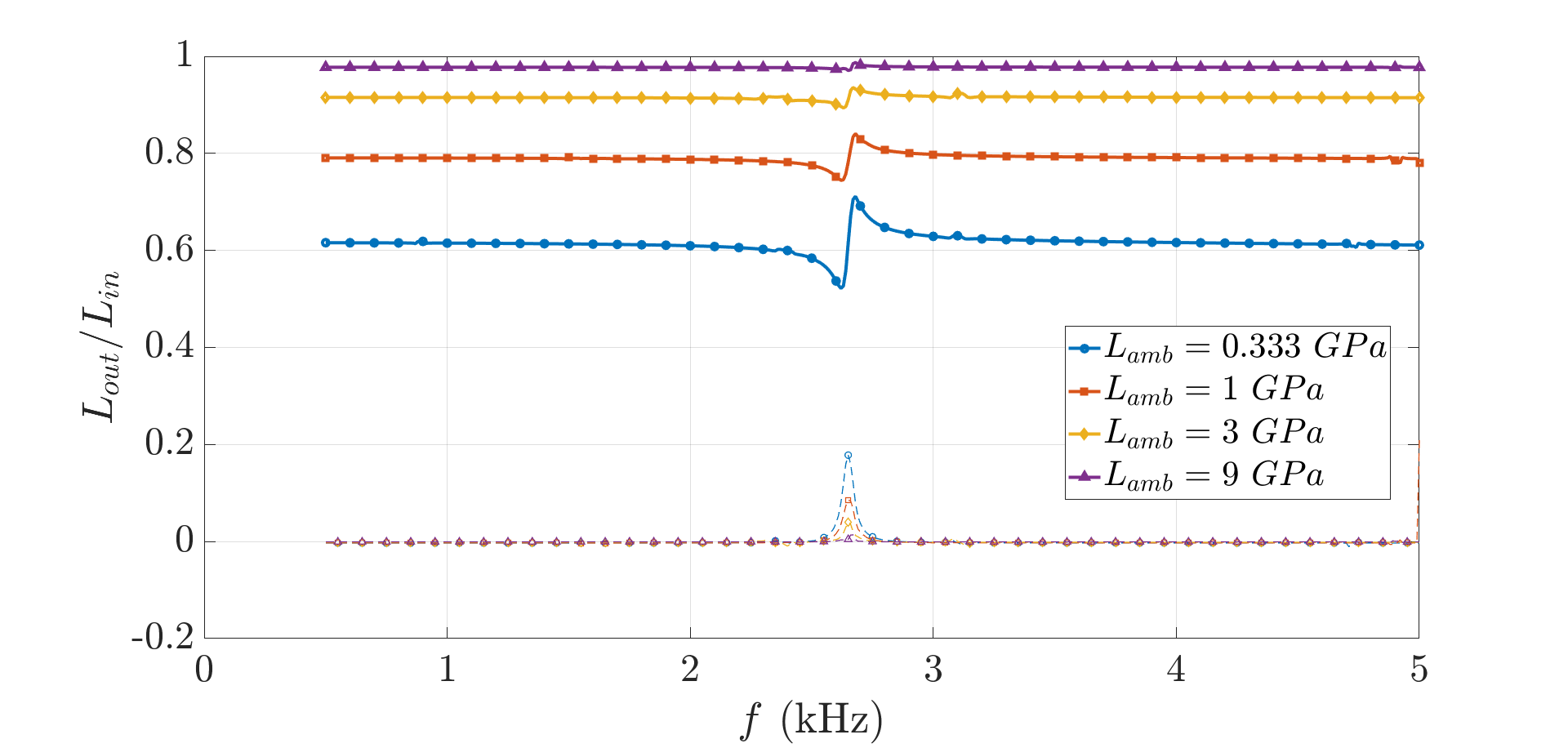}
			\caption{}
			\label{L_out_ratio}
		\end{subfigure}
		\begin{subfigure}{0.24\linewidth}
			\includegraphics[scale=0.15]{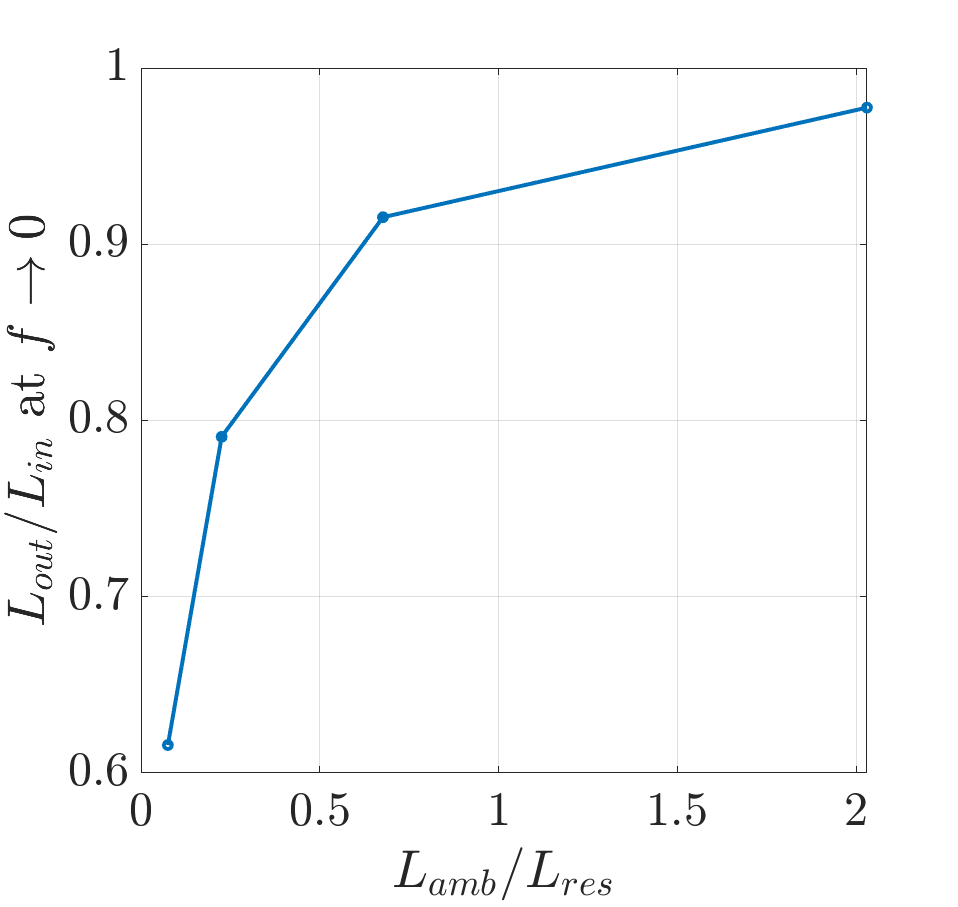}
			\caption{}
			\label{relationship} 
		\end{subfigure}
		\caption{(\subref{L_out_ratio}) The ratio of the skin to interior domain apparent modulus (simplified to its $f\rightarrow 0$ value in (\subref{relationship})) approaches a limit of 1 as $L_{amb}$ increases.}
		\label{L_ratio_relation} 
\end{figure}

The ratios of $Z_{out}/Z_{in}$ and $c_{out}/c_{in}$ remain relatively constant over the frequency range with a dependence on the ambient media modulus. In fact the ratio of densities appear to be nearly 1, independent of ambient properties. Figure~\ref{L_out_ratio} highlights the ratio of moduli for the skin to the interior domain, $L_{out}/L_{in}$, as the ambient media modulus changes. This is very useful for applying skin region effects to reduced order models or designing arrays with specific performance targets \cite{morris_expanding_2022}. Once the relationship between skin and interior domains is known for desired ambient media, the geometry of the outer cell can be adjusted accordingly. This relationship appears to have an asymptotic behavior as a function of $L_{amb}$ and $L_{res}$, evident in Figure~\ref{relationship}, but the exact physical description is yet to be determined. A closed form relationship or estimate may be developed with more analytical or numerical modelling, but is beyond the scope of the present work.  
%
%

\section{Phase Ambiguity, Unwrapping, and Mode Coupling} \label{branch_selection}
	
\begin{figure}[!ht]
	\centering
		\begin{subfigure}{0.35\linewidth}
			\includegraphics[scale=0.3]{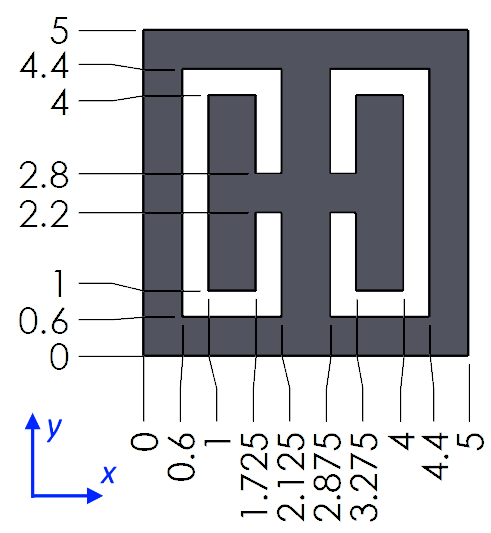}
			\caption{ }
			\label{geometry_38khzH}
		\end{subfigure}
		\begin{subfigure}{0.4\linewidth}
			\includegraphics[scale=0.25]{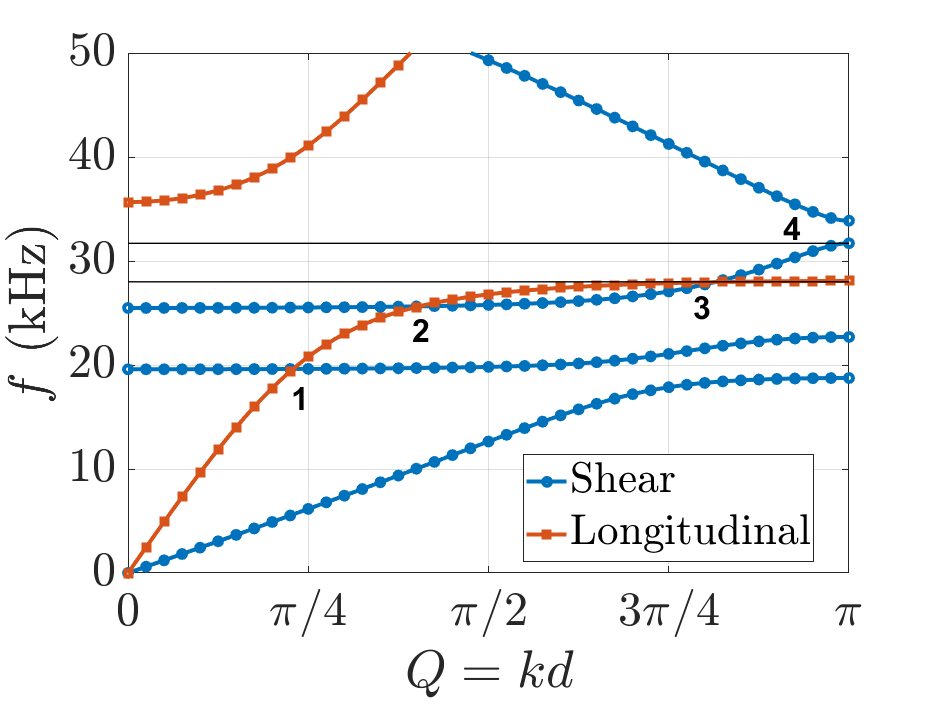}
			\caption{}
			\label{5H_band} 
		\end{subfigure}
		\caption{(\subref{geometry_38khzH}) The geometry of the \SI{5}{mm} H-design metamaterial cell. (\subref{5H_band}) The band structure calculated using eigenfrequency analysis with points of interest noted for discussion.}
		\label{5H_eigen} 
	\end{figure}

The unit cell analyzed in the previous sections has a particularly low longitudinal locally resonant frequency compared to its Bragg and shear modes, which avoids several potential complications. A second geometry was studied to demonstrate the potential for interference from modes other than the propagating longitudinal mode particularly near and within the stop band, where the transmission associated with longitudinal excitation is extremely weak for lossless and low loss media. A smaller unit cell made with the same resin (using loss factor $\eta_{res}=0.02$) is analyzed here for which a number of interfering features are intentionally present within and near its longitudinal stop band. This version of the H-design metamaterial with the cell size ($a_y = a_x = d/N$) reduced to \SI{5}{mm} has the band structure shown in Figure~\ref{5H_eigen}. The center of the \SI{8.6}{kHz} wide stop band is at \SI{32.4}{kHz}, achieving metamaterial performance within the ultrasonic range, with figure of merit $\lambda_{res}/a_x = 12.0$ at the center of the stop band.
The scattering results for a finite array are calculated with a few alterations compared to the \SI{10}{mm} cell version. The number of cells in the array is increased to $N=6$ to increase the transmission loss and highlight numerical issues that arise when transmission is especially low within the stop band. The ambient media material is assigned $L_{amb}=$ \SI{1}{GPa}, $\rho_{amb}=$\SI{1000}{kg/m^{3}}, and $\nu_{amb}=0.25$. The distances from the sample faces to the measurement locations have been changed to $x^{a,b}_{m1}=$ \SI{7.25}{cm} and $x^{a,b}_{m2}=$ \SI{7.75}{cm}, centered on a \SI{15}{cm} long ambient domain. The choice of measurement locations becomes more significant at higher frequencies when the wavelengths become comparable to the distance between $x^{a,b}_{m1}$ and $x^{a,b}_{m2}$. For each frequency $f$, the nodes for a standing wave will be located at intervals 
\begin{equation}
\frac{\lambda_{amb}}{2}=\frac{c_{amb}}{2f},  
\end{equation}
which will create discontinuities when the spacing of the measurements is equal to this spacing during the calculation of phase difference from Equation~\ref{gamma_matrix}. For this example, a spacing of $x^{a,b}_{m2}-x^{a,b}_{m1}=$ \SI{5}{mm} would introduce its first discontinuity at \SI{50}{kHz} which is just outside the range of interest. These disruptions can also be manipulated through the choice of ambient media ($c_{amb}$). 

\begin{figure}[!ht]
	    \centering
		\begin{subfigure}{0.4\linewidth}
			\includegraphics[scale=0.25]{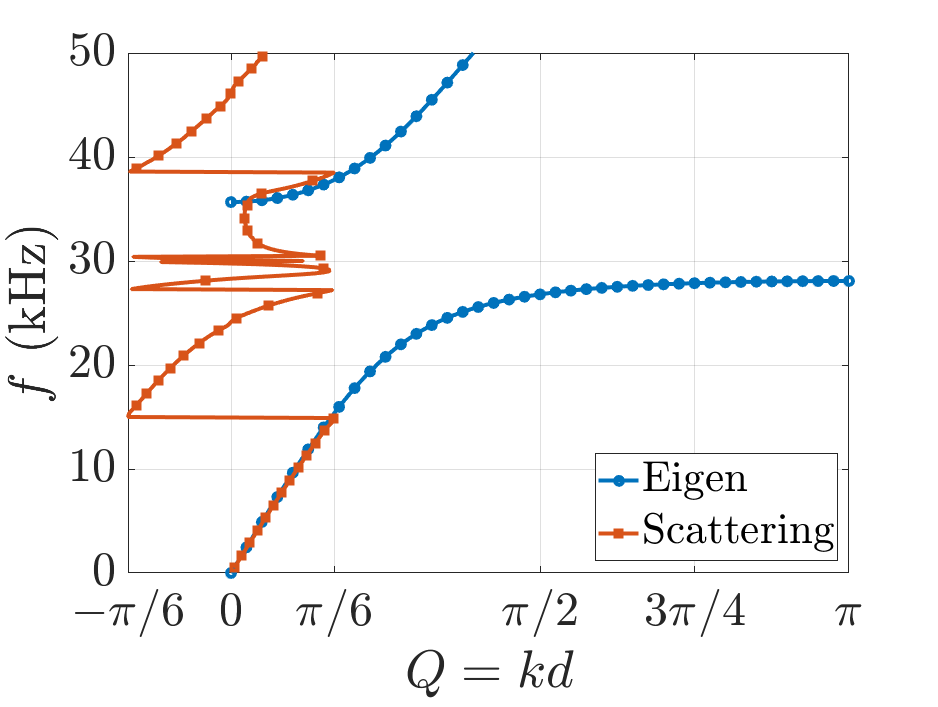}
			\caption{ }
			\label{branch_figa}
		\end{subfigure}
		\begin{subfigure}{0.4\linewidth}
			\includegraphics[scale=0.25]{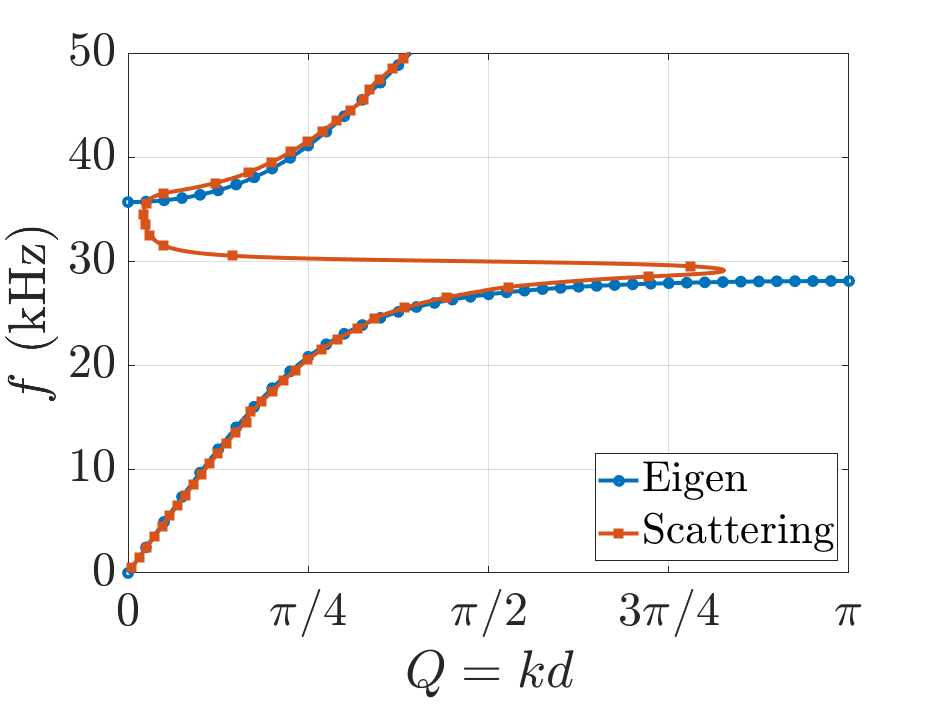}
			\caption{}
			\label{branch_figb} 
		\end{subfigure}
		\caption{(\subref{branch_figa}) The dispersion solution is bound between $-\pi/N$ and $\pi/N$ before branch corrections are applied. (\subref{branch_figb}) The branch corrected scattering dispersion curve is a close match to the infinite eigenfrequency solution.}
		\label{branch_fig} 
	\end{figure}

The immediate calculation of the the phase advance across the sample thickness ($kd$ from Equation~\ref{wave_vector}) provides solutions between $-\pi$ and $\pi$. This is clear when one seeks to reproduce the band structure with Equation~\ref{Q_eq}, where the longitudinal dispersion curve solution jumps from $-\pi/N$ to $+\pi/N$ multiple times if no branch corrections are performed. This behavior is shown in Figure~\ref{branch_figa}. To maintain the continuity of the phase as much as possible, an integer multiple of $2\pi$ may be added to $kd$ when a discontinuity is observed to recreate a continuous curve, a process also referred to as phase unwrapping: 
\begin{equation}
k = \frac{\log(e^{-ik d})}{-id}+n\frac{2\pi}{d}.
\label{wave_vector_shifted}
\end{equation}
Note that the results in this section do not separate the skin and interior regions, and therefore these quantities are functions of ambient properties and number of cells. However, the method described earlier has been applied to this micro-structure as well to independently determine the overall properties of skin region (functions of both unit cell and ambient material properties) and interior region (functions of only the unit cell geometry and properties) successfully. It is again observed that both sets are independent of sample thickness (number of cells). Moreover, it is also observed that the results closely follow those based on the eigenfrequency models through the first low frequency stop band and into the second pass band. The overall impedance begins to diverge from the eigensolution at higher frequencies, which is thought to be associated with limits of homogenization and will be studied further elsewhere.

\begin{figure}[!ht]
	    \centering
		\begin{subfigure}{0.4\linewidth}
			\includegraphics[scale=0.25]{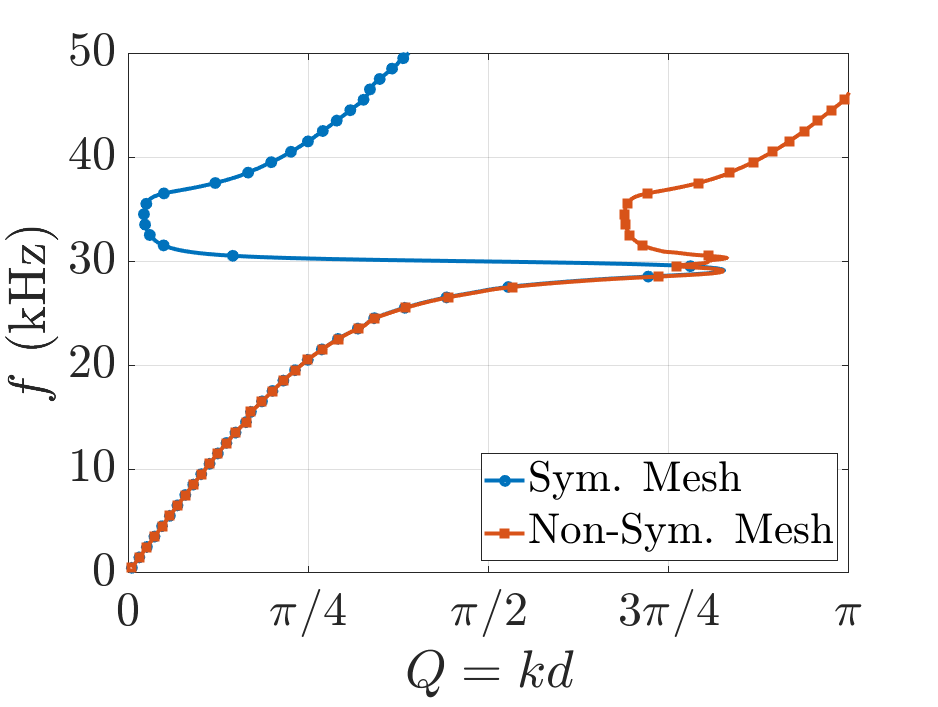}
			\caption{ }
			\label{sym_figa}
		\end{subfigure}
		\begin{subfigure}{0.4\linewidth}
			\includegraphics[scale=0.25]{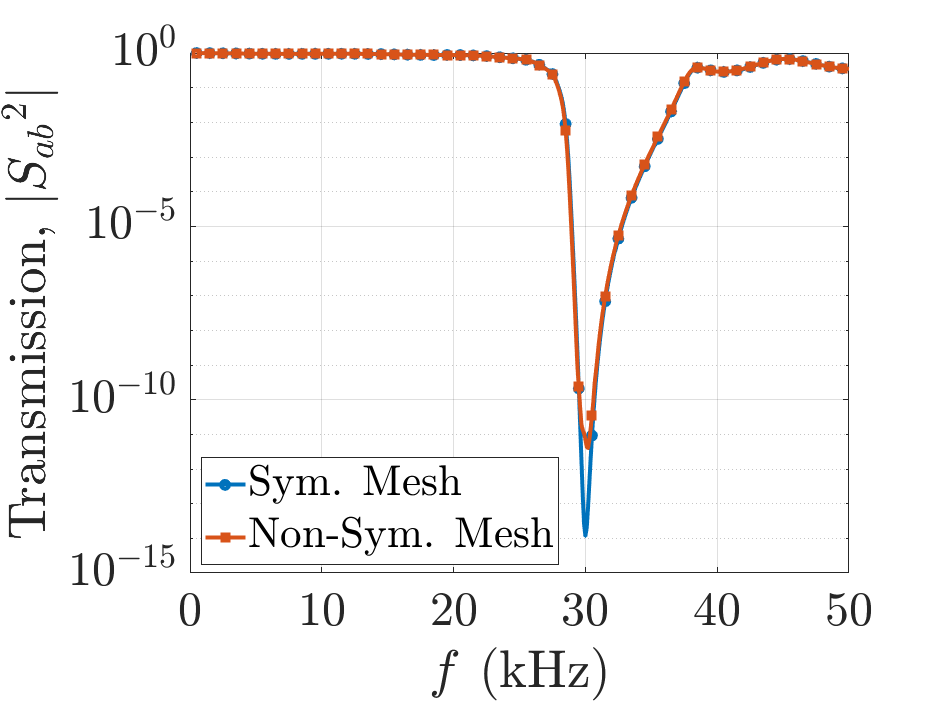}
			\caption{}
			\label{sym_figb} 
		\end{subfigure}
		\caption{Shear coupling in a non-symmetric mesh causes (\subref{sym_figa}) error in the longitudinal phase advance solution, evident around 30 kHz, as well as (\subref{sym_figb}) higher energy transmission that limits attenuation within the deepest region of the stop band.}
		\label{sym_fig} 
	\end{figure}

\begin{figure}[!ht]
		\centering
		\includegraphics [scale=0.25]{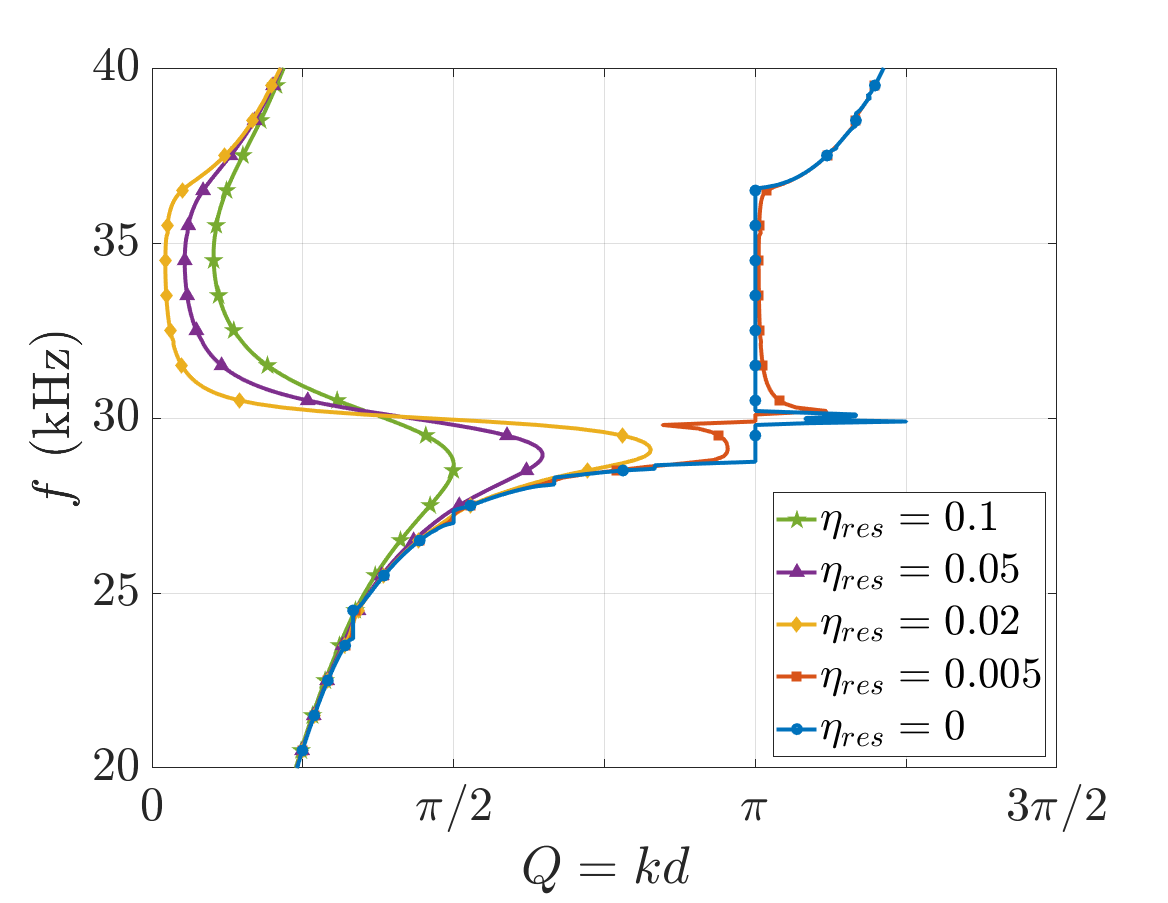}
		\caption{An example where loss limit technique become necessary, where very low loss $\eta_{res}$ constituent materials yield discontinuous solutions.}
		\label{38khz_loss} 
	\end{figure}

The transmission amplitude for the longitudinal stop band shown in Figure~\ref{branch_fig} is very close to zero around 30 kHz and a numerical issue could arise in certain cases. The computational mesh used for this figure is enforced to have similar symmetry as the unit cell. If the computational mesh is allowed to be asymmetric in a single cell (or in the presence of unavoidable symmetry-breaking imperfections in a physical sample), the enforcement of phase continuity will lead to different and unsatisfactory results; See Figure~\ref{sym_figa}. Continuity may however be restored, if transmission within the stop band is increased (either due to higher constituent loss or a smaller number of unit cells). 
The addition of loss and the limiting process (loss limit to zero) while maintaining the shape of the phase advance function has been used in parameter extraction technique to resolve phase-dependent shifts of the wave number.This resolves continuity enforcement challenges for low loss systems \cite{abedi_use_2020}. In practical applications, perfect symmetry will be almost impossible to enforce, so such techniques may be needed, but they may lead to unreliable parameter extractions within the stop band. An example of the application of loss limit approach is provided in Figure~\ref{38khz_loss} using a symmetric mesh, where the metamaterial constituent loss $\eta_{res}$ is varied. Even with mesh symmetry, as $\eta_{res}\rightarrow 0$ the solution for $Q$ becomes discontinuous. The higher loss cases can be used to inform the shape of the solution and the trend can be manually enforced. The interruption in $Q$ for Figures~\ref{sym_figa} and \ref{38khz_loss} both occur at 30 kHz. This feature also appears in the transmission curve shown in Figure~\ref{sym_figb}, where the non-symmetric mesh appears to lead to higher transmission than the symmetric mesh at the deeper regions of the stop band.
A closer examination of the cause of this discontinuity reveals an overlapping shear mode that may be activated with longitudinal excitation. Within the stop band and between 28 to \SI{32}{kHz}, a shear mode is present in the eigenfrequency determined band structure. It is observed that the discontinuous region identified in Figures~\ref{sym_fig} and \ref{38khz_loss} matches with the region in Figure~\ref{5H_band} between apparent crossing point 3 and the end of the shear pass band (point 4). Careful study of the crossing points (in particular point 3) and contrasting them in symmetric vs. asymmetric meshes indicate that in asymmetric meshes the crossing is avoided, instead demonstrating a level repulsion phenomenon \cite{mokhtari_scattering_2020,wang_exceptional_2022,amirkhizi_reduced_2018}. Around such points and due to the potential coupling of the two modes, phase continuity may not be successfully enforced. A further contributing factor is the different in energy flux features at point 3 for the two modes based on the energy velocity (presumably equal to group velocity) vs. points 1 and 2. In avoided crossings at points 1 and 2, the group velocity of the longitudinal dominant mode is finite and higher than that of the shear dominant modes, while at point 3 the opposite is true. This difference also contributes to the increased transfer of energy, via shear modes, but measured in the ambient domain as longitudinal waves, since coupling is present. In physical cases, unavoidable asymmetry is a cause for concern, however, higher non-zero losses generally limit the shear mode transmission. The practical performance is therefore likely to be far smoother, similar to the $\eta_{res}=0.02-0.1$ cases in Figure~\ref{38khz_loss}. This discussion provides the motivation behind the loss limit to zero approach. 

\section{Conclusions}

The presented multi-point scattering analysis procedure is a robust method for extracting the effective or apparent properties of frequency dependent finite metamaterial arrays. Thorough measurements at two points incoming and outgoing waves are decomposed independently for each ambient side. A study of this technique's sensitivity to modelling and physical parameters revealed several important considerations for expansion to a laboratory experiment. Two factors influencing the extracted parameters using this method include the outside media elastic modulus and the number of repeating unit cells across thickness (or total specimen thickness). A higher ambient domain modulus and higher number of unit cells yield extracted parameters closer to the result from infinite eigenfrequency analysis. The density of ambient media appears not to affect these results. Considering the array as a combination of skin (boundary or edge cells) and interior regions with associated differing overall properties appears to resolve these unwanted dependencies on the thickness and ambient media. The interior cells extracted properties are independent of the number of cells and ambient properties and they match closely with estimates based on infinite array eigenfrequency analysis, where such comparison is possible (i.e. outside the stop band and within the region). The comparison with and deviation from the eigenfrequency analysis estimates are subjects of a future study. However, skin region extracted parameters depend on the ambient media and dominate response when the array contains fewer cells. The ratio of impedance and wave number for skin vs interior cells is relatively constant and similar through the frequency range and may be modeled for application in design cases. In fact effective density of the skin and interior regions appears to be equal, therefore the distinction may be limited primarily to the effective moduli. When extracting the overall constitutive parameters, the continuity of the phase advance across the array (as a function of frequency) is utilized to unwrap the phase and determine phase velocity associated with the array. However, small asymmetries (even associated with computational mesh) may lead to coupling between modes and effectively limit the utility of the phase continuity assumption. This is particularly problematic when the transmission is very small, e.g. within the stop band of a very low loss system. In such cases, even minute coupling with other transmitting (e.g. shear) modes within the array, is measured in the outside domain as a net longitudinal transmission with complicated phase response. In general, separate measurements of coupled longitudinal and shear wave scattering will be required to determine the response of generally asymmetric cell arrays but care must be taken to handle such unwanted couplings when focusing on single mode behavior. For computational studies, loss limit to zero procedure may help overcome the discontinuity in phase advance estimates. The techniques expanded and evaluated in this paper provide a framework for experimental characterization of micro-structured media and suggests separation of the overall properties to skin (boundary or edge cells) and interior domains for a robust representation of such systems. 
	
\section*{Acknowledgements}

This work was supported by Cooperative Agreement W911NF-17-2-0173 between CCDEVCOM Army Research Laboratory and University of Massachusetts, Lowell.

\bibliographystyle{unsrt}
\bibliography{Scattering_Analysis_Journal}
	
\end{document}